\documentclass[letterpaper,11pt]{article}
\pdfoutput=1
\usepackage{jheppub}
\usepackage{slashed}
\usepackage{graphicx}
\usepackage{subfigure}
\usepackage{soul}
\usepackage{amsmath}
\usepackage{multirow}
\usepackage{tablefootnote}
\usepackage{hyperref}
\usepackage{epstopdf}
\usepackage{lscape}
\usepackage{url}
\usepackage[dvipsnames]{xcolor}

\newcommand{\beq}{\begin{equation}}
\newcommand{\eeq}{\end{equation}}
\newcommand{\bea}{\begin{eqnarray}}
\newcommand{\eea}{\end{eqnarray}}

\def\m1{M_1}
\def\m2{M_2}
\def\m3{M_3}

\def\ch10{\tilde \chi^0_1}

\def\non{\nonumber}

\def\gev{\,{\rm GeV}}

\def\to{\rightarrow}

\newcommand{\lsim}{\mathrel{\mathop{\kern 0pt \rlap
  {\raise.2ex\hbox{$<$}}}
  \lower.9ex\hbox{\kern-.190em $\sim$}}}
\newcommand{\gsim}{\mathrel{\mathop{\kern 0pt \rlap
  {\raise.2ex\hbox{$>$}}}
  \lower.9ex\hbox{\kern-.190em $\sim$}}}

\definecolor{pink}{RGB}{255,105,180}

\def\fbi{\,{\rm fb}^{-1}}

\def\figureautorefname~#1\null{Fig.\,#1\null}
\def\tableautorefname~#1\null{Tab.\,#1\null}

\def\equationautorefname~#1\null{Eq.\,(#1)\null}

\RequirePackage[normalem]{ulem}

\title{MSSM at future Higgs factories}

\author[\circ]{Honglei Li,}
\author[\star]{Huayang Song,}
\author[\star]{Shufang Su,}
\author[\#]{Wei Su,}
\author[\dagger,\ddagger]{Jin Min Yang}

\affiliation[\circ]{School of Physics and Technology, University of Jinan, Jinan, Shandong 250022, China }
\affiliation[\star]{Department of Physics, University of Arizona, Tucson, Arizona 85721, USA}

\affiliation[\#]{ARC Centre of Excellence for Dark Matter Particle Physics and CSSM,
Department of Physics, University of Adelaide, SA 5005, Australia}
\affiliation[\dagger]{CAS Key Laboratory of Theoretical Physics, Institute of Theoretical Physics, \\
Chinese Academy of Sciences, Beijing 100190, China.}
\affiliation[\ddagger]{School of Physical Sciences, University of Chinese Academy of Sciences, Beijing 100049, China}

\emailAdd{sps\_lihl@ujn.edu.cn, huayangs@email.arizona.edu, shufang@arizona.edu, wei.su@adelaide.edu.au, jmyang@itp.ac.cn}

\abstract{In this work, we study the implication of Higgs precision measurements at future Higgs factories on the MSSM parameter space, focusing on the dominant stop sector contributions.  We perform a multi-variable fit to both the signal strength for various Higgs decay channels at Higgs factories and the Higgs mass. The $\chi^2$ fit results show sensitivity to $m_A$, $\tan\beta$,  stop mass parameter $m_{\rm SUSY}$ as well as the stop left-right mixing parameter $X_t$. We also study the impact of the Higgs mass prediction on the MSSM and compare the sensitivities of different Higgs factories. }


\preprint{ADP-20-30/T1140
}

\begin{document}
\maketitle
\flushbottom

\section{Introduction}

After the discovery of a 125 GeV SM-like Higgs boson at the LHC~\cite{Aad:2012tfa,Chatrchyan:2012ufa},  several proposals of a Higgs factory have been considered for precise measurements of the Higgs mass and couplings, including the Circular Electron Positron Collider (CEPC) in China~\cite{CEPCStudyGroup:2018ghi}, the electron-positron stage of the Future Circular Collider (FCC-ee) at CERN~\cite{Mangano:2018mur,Benedikt:2018qee}, and the International Linear Collider (ILC) in Japan~\cite{Bambade:2019fyw}.  With about $10^6$ Higgs produced, the Higgs mass can be measured to an accuracy of about 10 MeV. Various Higgs couplings to the Standard Model (SM) particles can be measured at about 1\% level, while $hZZ$ coupling can be measured at about 0.2\%.  If no deviation from the SM predictions is observed at future Higgs factories, severe constraints can be imposed on the parameter space of new physics models. On the other hand, if certain deviations are observed, it will provide great insights into new physics models beyond the SM, in particular, on the masses and couplings of new particles, which can be cross checked with direct searches of new particles at future high energy colliders. 

In this work we focus on the implication of Higgs precision measurements on the  Minimal Supersymmetric Standard Model (MSSM), which is one of the well motivated new physics models proposed to solve the hierarchy problem~\cite{Chang:2005ht,Martin:1997ns}. The Higgs sector of the MSSM is the same as that of the Type-II two Higgs doublet model (2HDM), with one Higgs doublet $H_u$ coupling to the up type quarks, and the other Higgs doublet $H_d$ coupling to down type quarks and charged leptons.  After the electroweak symmetry breaking, there are five physical Higgses:  two CP-even Higgses $h$, $H$, one CP-odd Higgs $A$ and a pair of charged ones $H^{\pm}$. In our analyses below, we take the light CP-even Higgs $h$ to be the observed 125 GeV SM-like Higgs\footnote{The other possibility of the heavy CP-even Higgs $H$ being the 125 GeV Higgs is tightly constrained by the existing experimental searches, as shown in Ref.~\cite{Christensen:2012ei}.}.    At tree level, the masses of MSSM Higgses   are completely determined by only two parameters: the CP-odd Higgs mass $m_A$ and the ratio of two Higgs vacuum expectation values $\tan\beta$.  The mass of the SM-like light CP-even Higgs, however, receives large radiative corrections, with the dominant   contribution from the supersymmetric top partners.  The Higgs couplings to the SM sector also receive radiative corrections, characterised by the effective mixing angle $\alpha_{eff}$.   In addition, the bottom Yukawa coupling   receives large vertex corrections.  Furthermore, Higgs couplings to a pair of photons or gluons receive loop contributions from the stop sector, which are at about the same order as the SM contributions.

While in general there are more than 100 parameters in the MSSM, when we focus on the Higgs sector and the dominant loop contributions from the stop sector, only four are the most relevant: $\tan\beta$, $m_A$, soft supersymmetry (SUSY)  breaking stop mass parameter $m_{\rm SUSY}$\footnote{For simplicity, we have taken the left and right stop mass parameters to be the same.}, and the left-right stop mixing parameter $X_t$.   Other parameters, such as the mass parameters for the sbottom and gluinos could enter as well.  Those effects become important in a particular corner of the parameter space, which is left for future dedicated studies.

 To study the implication of Higgs precision measurements on the parameter space of the MSSM, we perform a multi-variable $\chi^2$ fit to both the signal strength $\mu$ for various Higgs decay channels and the Higgs mass.  Earlier works on the implication of Higgs precision measurements mostly focused on the loop induced channels $h \to gg$ and $h \to \gamma\gamma$~\cite{Fan:2014axa,Fan:2014txa,Drozd:2015kva,Drozd:2015rsp}, given that both the SM and the MSSM contributions enter at the same order. In our work, we include all the Higgs decay channels measured at Higgs factories, as well as the Higgs mass.  The MSSM predictions of those quantities have been studied extensively in the literature~\cite{Carena:1995bx, Heinemeyer:1998yj,Carena:2002qg}.    For the MSSM corrections to the Higgs couplings to the SM particles, we adopt the $\alpha_{\rm eff}$ method~\cite{Dabelstein:1995js,Carena:1995bx}. We also include the additional vertex corrections to the bottom Yukawa and loop induced couplings of $hgg$ and $h\gamma\gamma$.  We use the state-of-art program FeynHiggs~\cite{Bahl:2018qog, Bahl:2017aev, Bahl:2016brp, Hahn:2013ria, Frank:2006yh, Degrassi:2002fi, Heinemeyer:1998np, Heinemeyer:1998yj} to obtain $\alpha_{\rm eff}$ and $m_h$ in the framework of the MSSM.

In Sec.~\ref{sec:higgs_precision}, we briefly summarize the Higgs precision measurements at various Higgs factories.  We also introduce the $\chi^2$ fit formalism used in our analyses.  In Sec.~\ref{sec:mssm}, we discuss the MSSM Higgs sector and stop sector that are needed in our analyses, as well as the SM-like Higgs couplings in the MSSM.  In Sec.~\ref{sec:direct}, we summarize the current direct search limits on the mass of the CP-odd Higgs and the stop sector.  In Sec.~\ref{sec:chi2}, we perform  detailed analyses of various contributions to the total $\chi^2$.  In Sec.~\ref{sec:global}, we present the 95\% C.L. allowed region of the MSSM parameter space under the CEPC precisions.  In Sec.~\ref{sec:comparison}, we compare the reach of the CEPC, the FCC-ee and the ILC.   We reserve Sec.~\ref{sec:conclusion} for conclusions.

\section{The Higgs precision measurements and $\chi^2$ fit}
\label{sec:higgs_precision}

The analyses of precision measurements of Higgs decay channels have been performed at the CEPC~\cite{CEPC-SPPCStudyGroup:2015csa,CEPCStudyGroup:2018ghi}, the FCC-ee~\cite{Abada:2019lih,Abada:2019zxq,Gomez-Ceballos:2013zzn,Blondel:2019yqr}, as well as the ILC~\cite{Baer:2013cma,Bambade:2019fyw,Fujii:2019zll,Fujii:2020pxe} in recent years.  A summary of the most updated results on   $\Delta(\sigma\times{\rm Br})/(\sigma\times{\rm Br})$, as well as the total production cross section $\Delta \sigma/\sigma$, can be found in Table 3 in Ref.~\cite{Chen:2019pkq}, which will be used in our current study.   The dominant production channel at 240$-$250 GeV is the associated  $Zh$ production, with   the best measured channel being $h\rightarrow b\bar{b}$, given its large decay branching fraction. A precision of about 0.3\% can be achieved for this channel.
The precisions for $h\rightarrow gg,\ WW^*, \tau^+\tau^-$ are about 1\%, while $h\rightarrow c\bar{c}$ is about 2$-$3\%.  The precisions for $h\rightarrow ZZ^*, \gamma\gamma$ are worse, about 5$-$7\% given its suppressed decay branching fractions.     The sensitivities for the  three Higgs factories are very similar. Weak boson fusion (WBF) process $e^+e^-\rightarrow \nu \bar\nu h$ becomes more important at higher center of mass energy, with a precision of about 0.23\% can be achieved for $h\to b\bar{b}$ channel at the ILC 500 GeV with 4 ${\rm ab}^{-1}$ integrated luminosity~\cite{Bambade:2019fyw,Fujii:2019zll}.

To analyze the implication of Higgs precision measurements on the MSSM parameter space, we perform a multi-variable $\chi^2$ fit
\begin{equation} 
  \chi^2_{total} = \chi^2_{m_h} + \chi^2_{\mu} = \frac{(m_h^{\rm MSSM}-m_h^{\rm obs})^2}{(\Delta m_h)^2} + \sum_{i=f,V..} \frac{(\mu_i^{\rm MSSM}-\mu_i^{\rm obs})^2}{(\Delta \mu_i)^2},
  \label{eq:chi2}
\end{equation}
in which $\mu_i^{\rm{MSSM}}=(\sigma\times\textrm{Br}_i)_{\rm{MSSM}}/(\sigma\times\textrm{Br}_i)_{\rm{SM}}$ is the signal strength for various Higgs search channels. Here $\chi^2_{m_h}$ and $\chi^2_{\mu}$ refer to contributions to the overall $\chi^2_{total}$  from the Higgs mass and signal strength measurements, respectively.  For $\chi^2_{m_h}$, given the small experimental uncertainties and the relatively large theoretical uncertainties in determining   $m_h$ in the MSSM, we set $\Delta m_h$ to be 3 GeV, taking into account uncertainties coming from higher order radiative corrections ~\cite{Degrassi:2002fi,Frank:2006yh,Hahn:2013ria,Bahl:2016brp}, as well as propagating uncertainties from SM input parameters like $m_t$.   Results with smaller $\Delta m_h= 1$ GeV and 2 GeV are also presented in Sec.~\ref{sec:global} to show the impact of possible future improvement in $m_h$ calculation including higher order corrections~\cite{Bahl:2019hmm}.  For $\chi^2_\mu$, $\Delta \mu_i$   is  the experimental expected precision in determining the signal strength for a particular Higgs decay channel.  

For future Higgs factories, $\mu_i^{\rm{obs}}$ are set to be unity in our  analyses, assuming no deviations from the SM predictions are observed\footnote{If deviations are observed in the future, we can use the same $\chi^2$ fit method to determine the constrained parameter space, with $\mu_i^{\rm{obs}}$ being the observed experimental central value~\cite{Han:2020lta}.  }.  Usually, the correlations among different search channels at Higgs factories are not provided and are thus assumed to be zero.

 In our analyses, we determine the allowed parameter region at the $95\%$ Confidence Level (C.L.) by a multi-variable fit to the Higgs decay signal strengths of various channels and Higgs mass.   For the one-, two- or three-parameter fit, the corresponding $\Delta\chi^2=\chi^2-\chi_{\rm{min}}^2$ at 95\% C.L. is 3.84, 5.99 or 7.82, respectively. Note that when we present our results of three-parameter fit in Sec.~\ref{sec:global}, we project the three dimensional space onto two-dimensional plane for several benchmark points in the third dimension of the parameter space.  Most of the results presented below are for the CEPC precisions.  We  compare the reaches of the CEPC, the FCC-ee and the ILC in Sec.~\ref{sec:comparison}.

\section{The Higgs and stop sector of the MSSM}
\label{sec:mssm}

\subsection{The Higgs mass in the MSSM}
In our analyses, we identify the light CP-even Higgs $h$ in the MSSM as the observed 125 GeV SM-like Higgs.  Its mass receives large radiative corrections, dominantly from the stop sector, as well as the sbottom sector at large $\tan\beta$. There have been extensive studies of the MSSM loop correction to the Higgs masses up to next to next order~\cite{Carena:1995bx, Heinemeyer:1998yj,Carena:2002qg},   which includes full one-loop contributions as well as the leading two-loop contributions $\mathcal{O}\left(\alpha_{t} \alpha_{s}, \alpha_{b} \alpha_{s}, \alpha_{t}^{2}, \alpha_{t} \alpha_{b}, \alpha_{b}^{2}\right)$ to the Higgs
two-point functions.   There are also works considering the three-loop effects at order $\mathcal{O}(\alpha_{t, b} \alpha_{s}^{2}, \alpha_{t, b}^{2} \alpha_{s}, \alpha_{t, b}^{3})$~\cite{Stockinger:2018oxe}, as well as approximate evaluation at   order $\mathcal{O}(\alpha_{t}^2 \alpha_{s}^{2})$~\cite{Harlander:2018yhj}.

The CP-even Higgs mass matrix is given by 
\begin{equation}
 \cal{M}_{\rm Higgs} \rm = \frac{ \sin 2 \beta }{2}\left( \begin{array}{ll}
     \cot \beta \ m_Z^2 + \tan \beta \ m_A^2 & ~~~~- m_Z^2 - m_A^2\\
     - m_Z^2 - m_A^2  & \tan \beta \ m_Z^2 + \cot \beta \ m_A^2
     \end{array} \right) +
     \left( \begin{array}{ll}
     \Delta_{11} & \Delta_{12}
     \\
     \Delta_{12} & \Delta_{22}
     \end{array}
     \right),
\label{eq:glalphaap}
\end{equation}
with the first term being the tree-level contributions and   $\Delta_{11},\Delta_{12},\Delta_{22}$ in the second term are the loop-induced Higgs mass corrections~\cite{Dabelstein:1995js,Carena:1995bx,Harlander:2017kuc}.  The  masses for the CP-even Higgses are obtained by the diagonalization of the mass matrix: 
\begin{eqnarray}
 M^2_{H,h,\, eff} & = & \frac{m_A^2 + m_Z^2 + \Delta_{22} + \Delta_{11}}{2} \pm \Big(
 \frac{ (m_A^2 + m_Z^2)^2 + ( \Delta_{22} - \Delta_{11} )^2}{4} - m_A^2 m_Z^2 \cos^2 2\beta \nonumber \\
 &&  +\frac{( \Delta_{22}- \Delta_{11}) \cos 2\beta}{2}  (m_A^2 - m_Z^2)
 - \frac{\Delta_{12} \sin 2\beta}{2} (m_A^2 + m_Z^2) + \frac{\Delta_{12}^2}{4} \Big)^{1/2}.
\label{eq:hmass_eff}
\end{eqnarray}
The   effective mixing angle $\alpha_{eff}$ between CP-even scalars is defined by
\begin{equation}
\left(\begin{matrix}h\\H\end{matrix}\right)=\left(\begin{matrix}
\cos\alpha_{eff} & \sin\alpha_{eff}\\
-\sin\alpha_{eff} & \cos\alpha_{eff}
\end{matrix}\right)\left(\begin{matrix}
\text{Re}H_u^0-v_u \\ \text{Re}H_d^0-v_d
\end{matrix}\right),
\end{equation}
which takes the form of
\begin{eqnarray}\label{eq:alp_eff}
  \tan \alpha_{eff} &=& \frac{-(m_A^2+m_Z^2)\sin\beta \cos \beta + \Delta_{12}}{m_Z^2 \cos^2 \beta +m_A^2 \sin^2 \beta +\Delta_{11}-m_{h^0,eff}^2}.
\end{eqnarray}

Out of all the supersymmetric particles, the stop sector gives the dominant loop contributions to the Higgs sector.  The stop mass matrix depends on the $H_u -H_d$ mixing parameter $\mu$ and  soft SUSY breaking parameters $m_{\tilde{Q}}$, $m_{\tilde{t}_R}$, and trilinear coupling $A_t$:
\begin{equation}
 \mathcal{M}_{\rm \tilde{t}}^{\rm 2} \rm = \left( \begin{array}{ll}
 m_{\tilde{Q}}^2 + m_t^2 + m_Z^2 (\frac{1}{2} - \frac{2}{3} s_W^2) \cos 2 \beta &
 ~~~~~m_t (A_t- \mu  \cot \beta  ) \\
 m_t (A_t - \mu  \cot \beta  ) &
 m_{\tilde t_R}^2 + m_t^2 +\frac{2}{3} m_Z^2  s_W^2 \cos 2 \beta
 \end{array} \right).
\label{eq:stmatrix}
\end{equation}
The stop left-right mixing parameter is defined as $X_t \equiv A_t - \mu  \cot \beta$, which enters the off-diagonal term, and plays an important role in the radiative corrections to the Higgs mass.    For our analyses below, we assume  mass degeneracy of left- and right-handed top squarks and take the most relevant model   parameters as:
\begin{equation}
\tan \beta, m_A,  m_{\rm SUSY}\equiv m_{\tilde Q} = m_{\tilde t_R}, X_t.
\label{eq:free_para}
\end{equation}

\subsection{Higgs couplings with $\alpha_{eff}$ method}
 
The effective lagrangian of the Higgs couplings to pair of fermions and gauge bosons can be written as~\cite{Henning:2014wua}   
\begin{eqnarray}\label{eq:eff_kappa}
\mathcal{L}=&& \kappa_Z \frac{m_Z^2}{v}Z_{\mu}Z^{\mu}h+\kappa_W \frac{2 m_W^2}{v}W_{\mu}^+ W^{\mu-}h + \kappa_g \frac{\alpha_s}{12 \pi v} G^a_{\mu\nu}G^{a\mu\nu}h + \kappa_{\gamma} \frac{\alpha}{2\pi v} A_{\mu\nu}A^{\mu\nu}h \nonumber \\
&&  -\Big( \kappa_t \sum_{f=u,c,t} \frac{m_f}{v}f \bar f + \kappa_b \sum_{f=d,s,b} \frac{m_f}{v}f \bar f + \kappa_{\tau} \sum_{f=e,\mu,\tau} \frac{m_f}{v}f \bar f \Big)h
\label{eq:kappa}
\end{eqnarray}
with $\kappa_i = \frac{g_{hii}^{BSM}}{g_{hii}^{SM}}$ being the Higgs coupling normalized to the SM value.  Given that the Yukawa coupling structure of the MSSM is the same as that of the Type-II 2HDM, $\kappa_u$, $\kappa_{d,l}$ and $\kappa_V$ follow the tree-level expression of the Type-II 2HDM, with the mixing angle $\alpha$ being replaced by the effective mixing angle $\alpha_{eff}$~\cite{Dabelstein:1995js,Carena:1995bx}, including radiative corrections: 
\begin{eqnarray}
  k_u = \frac{\cos \alpha_{eff}}{\sin \beta} ,
  k_{d,l} = -\frac{\sin \alpha_{eff}}{\cos \beta},
   k_V= \sin (\beta-\alpha_{eff}).
   \label{eq:kappaf}
\end{eqnarray}
This is the so-called ``$\alpha_{eff}$ method"~\cite{Dabelstein:1994hb}, which is used 
in our analyses to count for the MSSM loop corrections to the SM-like Higgs couplings to the SM particles.   This effective method is in good agreement to the full loop results~\cite{Dabelstein:1995js, Heinemeyer:2000fa}, under the heavy gluino mass assumption that we adopted in our analyses. 

Given the high experimental precision in $h\to b\bar{b}$ channel: $\Delta \mu_{b}=0.27\%$   at the CEPC~\cite{CEPCStudyGroup:2018ghi,Gu:2017ckc}, and large ${\rm Br}_{h\to b\bar{b}}=57.7\%$,   Higgs factories are particularly sensitive to MSSM contributions to $\kappa_b$.   In addition to the loop contributions to $\alpha_{eff}$, which enters $\kappa_b$ via Eq.~(\ref{eq:kappaf}), additional MSSM loop corrections to $\kappa_b$ are included in our analyses, which is characterized by $\Delta m_b$. 
\begin{equation}
\kappa_b =  -\frac{\sin \alpha_{eff}}{\cos\beta} \tilde \kappa^{b}_h,\ \ \ \tilde \kappa^{b}_h=  \frac{1}{1+\Delta m_b}\left(1-\Delta m_b \frac{1}{\tan \alpha_{eff} \tan \beta}\right).
\label{eq:kb_stop}
\end{equation}
Assuming large sbottom and gluino masses, the dominant loop contribution to $\Delta m_b$ comes from the stop sector~\cite{Carena:1999py}: 
\begin{equation}
\label{eq:dmb_stop}
\Delta m_b^{\rm stop} = \frac{h_t^2}{16 \pi^2}\mu A_t \tan \beta I(m_{\tilde t_1}, m_{\tilde t_2}, \mu).
\end{equation}

The loop-induced Higgs couplings, $hgg$ and $h\gamma\gamma$ receive contributions from the SUSY sector as well, which are of the same order as the SM contributions.  Therefore, $hgg$ and $h\gamma\gamma$ could provide extra sensitivity to the MSSM parameter space~\cite{Fan:2014axa,Drozd:2015kva}.  In particular, given the experimental precision for $hgg$ channel is about 1\% at Higgs factories, this channel is particularly sensitive to stops running in the loop.   Contributions from the sbottom sector 
are typically at least an order of magnitude smaller than those from the stop sector, even in the case of large $\tan\beta$~\cite{Drozd:2015kva}.   To focus on the dominated effects,   we   do not include the sbottom effects in our analyses.   

The signal strength $\mu_i$ that enters the $\chi^2$ analyses includes the MSSM contributions to both the Higgs production and decays.  We use the state-of-art program FeynHiggs~\cite{Heinemeyer:2004ms,Bahl:2016brp,Bahl:2018qog,Bahl:2019hmm} to obtain $m_h$, $\alpha_{eff}$ and $\Delta m_b$, calculating the various $\kappa$s as defined in Eq.~(\ref{eq:kappa}), which are fed into the evaluation of signal strength $\mu_i^{\rm MSSM}$.

\section{Direct search limits from the LHC}
\label{sec:direct}

 Other than the studies for the SM-like Higgs, there have been extensive searches for MSSM heavy Higgses at the LHC.  Given the light CP-even Higgs as the observed 125 GeV SM-like Higgs,  scenarios such as $m_h^{mod}$~\cite{Carena:2013qia}, $M_h^{125}$~\cite{Bahl:2018zmf} and hMSSM~\cite{Djouadi:2015jea} are proposed to test the model parameter spaces in the $m_A-\tan\beta$ plane.  Based on the data collected during the LHC Run 2 with an integrated luminosity of 139 $\fbi$ at $\sqrt{s}=13$ TeV,  the ATLAS collaboration searched for the heavy neutral Higgs bosons over the mass region 0.2$-$2.5 TeV with   $A/H \to \tau^+\tau^-$ decay~\cite{Aad:2020zxo}.  In the $M_h^{125}$ scenario, the data exclude the parameter space of $\tan\beta>8$ for $m_A= 1.0$ TeV,   $\tan\beta>21$ for $m_A= 1.5$ TeV, and $\tan\beta>60$ for $m_A=2.0$ TeV, which are the strongest exclusion limits in the large $\tan\beta$ region.
 Exclusion from $A/H \to b \bar {b}$ is weaker: for $\tan\beta$ between $20-60$, $m_A$ in the mass region of 0.45$-$0.9 TeV has been excluded with  $bbH/A$ production in the 
 scenarios of hMSSM~\cite{Aad:2019zwb}.     Results from CMS are similar~\cite{Sirunyan:2018zut}.

 In the low $\tan\beta$ region, $bb$ and $\tau\tau$ channels are less constraining given the reduced Yukawa couplings.  CMS searches with $A/H \to t \bar t$ exclude the value of $m_{A}$ at 400 (700 GeV) for $\tan\beta$ below 1.5 (1.0) ~\cite{CMS:2019lei}.  Decay modes of $H\to ZZ$, $A\to hZ$, $H\to WW$, and $H\to hh$ also constrain the parameter space in the low $\tan\beta$ region.  Combining the results from these channels, the mass region of 200$-$600 GeV is excluded with $\tan\beta$ value between 1 to 6  at both the ATLAS and CMS experiments~\cite{Aaboud:2017rel,Aaboud:2017cxo,Aaboud:2017gsl,Aad:2019uzh,Sirunyan:2019pqw,Sirunyan:2017djm}.   In addition, CMS searches of  $A\to hZ \to \tau\tau\ell\ell$ exclude $\tan\beta$ values below 1.6 at $m_A=220$ GeV and 3.7 at $m_A=300$ GeV~\cite{Sirunyan:2019xjg} in the hMSSM scenario.  

Searches for charged Higgses produced either in the top quark decay (for $m_H^\pm < m_t$) or in associated with a top quark (for $m_H^\pm > m_t$), with the subsequently decay of $H^+\to \tau  \nu$, are performed in the context of hMSSM at the LHC. For light charged Higgs with $m_{H^\pm}<m_t$,   $m_{H+}\le 160$ GeV is excluded~\cite{Aaboud:2018gjj}.    For heavy charged Higgs, the region of  $\tan\beta=20 - 60$  is excluded with $m_{H^\pm}$ from 200  to 1100 GeV~\cite{Aaboud:2018gjj}.   $H^+\to t b$ decay mode is sensitive to the  low $\tan\beta$ region.   Value of $\tan\beta =1.5 - 0.4$ are excluded in the $m_{H^+}$ range of 200 GeV to 1.5 TeV in the context of $m_h^{mod}$ scenario~\cite{Sirunyan:2019arl, Aaboud:2018cwk}.    

 For the stop sector, the limits are more complicated given their dependence on the mass spectrum of charginos and neutralinos, as well as the corresponding decay branching fractions. Several channels of the stop decay  to the lighter superparticles have been explored.  For  $\tilde{t}_1 \to t \tilde{\chi} _1^0 /bW\tilde{\chi} _1^0/bff^\prime\tilde{\chi} _1^0$, the latest results show that the stop  mass region of  $ m_{\tilde{t}_1}<1.2$ TeV is excluded for $\tilde{\chi} _1^0$ mass below about 500 GeV~\cite{Aad:2020sgw, Sirunyan:2019glc}.     With a light slepton, $\tilde{t}_1 \to b \tilde{\chi} _1^{+} \to b\nu  \tilde{\ell} \to b \nu \ell \tilde{\chi} _1^0 $ decay channel can exclude masses up to about 1.4 TeV for $\tilde{t}_1$ and 900 GeV for $\tilde{\chi} _1^0$ with $m_{\tilde{\chi} _1^{0}}<m_{\tilde{\ell}}<m_{\tilde{\chi} _1^{+}}$~\cite{Sirunyan:2020tyy}.  
  
 
\section{Contributions to $\chi^2$} 
\label{sec:chi2}
\begin{figure}[t]
\begin{center}
\includegraphics[width=15cm]{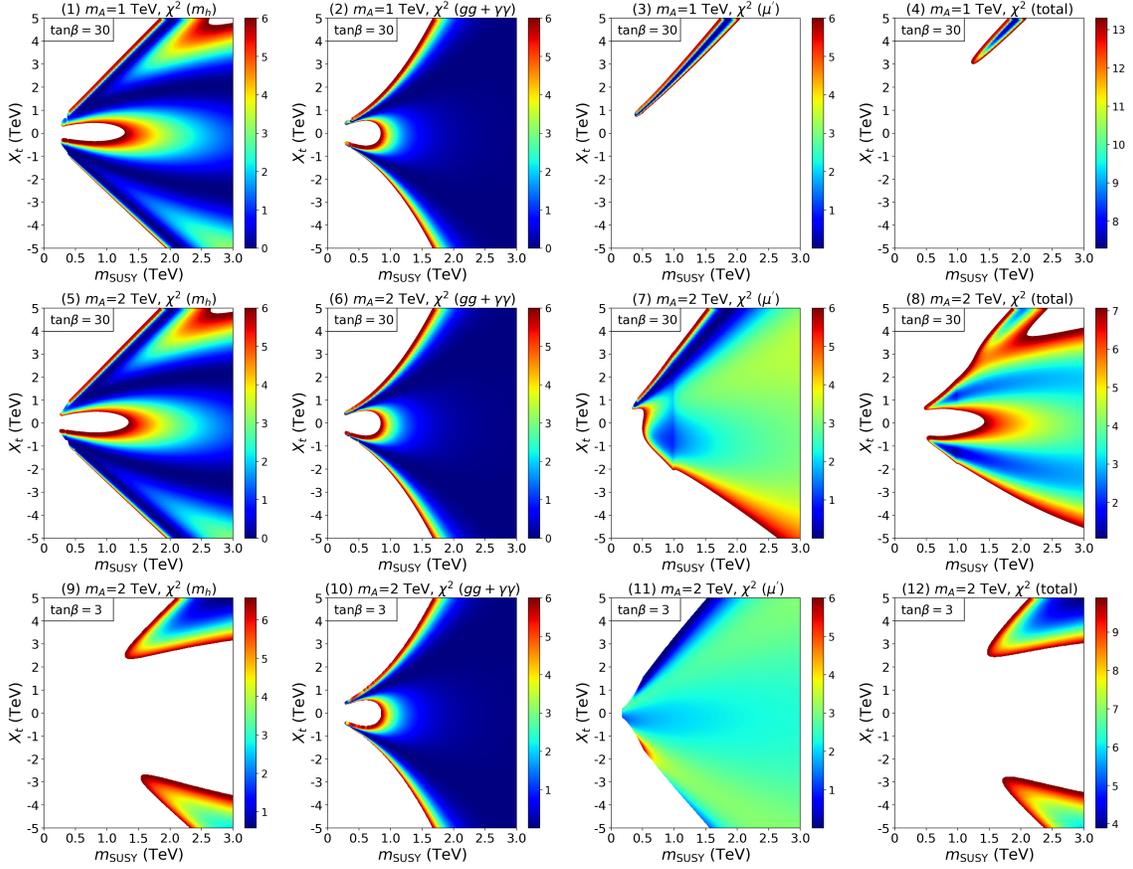}
\caption{95\% C.L.  allowed region in the plane of $m_{\rm SUSY} - X_t$ under CEPC precisions for  $(m_A,\tan\beta)=(1\ {\rm TeV}, 30)$ (upper panels), (2 TeV, 30) (middle panels) and (2 TeV, 3) (lower panels).  For each row, the panels from left to right show $\chi^2_{m_h}$, $\chi^2_{gg+\gamma\gamma}$, $\chi^2_{\mu^\prime}$, and $\chi^2_{total}$.  Different color band corresponds to the $\chi^2$ value.  See text for details.}
\label{fig:atmsf_analysis}
\end{center}
\end{figure}

In this section, we choose several typical sets of MSSM model parameters to study the various contributions to $\chi^2_{total}$: 
\begin{eqnarray}\label{eq:para_ana}
  m_A = 1000/2000 \text{~GeV}, \mu = 500 \text{~GeV}, \tan \beta= 3/30,\\ \non
   X_t \in (-5000, 5000) \text{~GeV},  m_{\rm SUSY} \in (200, 3000) \text{~GeV}.
\end{eqnarray}
To identify the stop contributions, we decouple the mass of  other sfermions and gluinos. 
We scan over the parameter space  of $X_t$ and $m_{\rm SUSY}$ to explore the various contributions to the overall $\chi^2_{total}$   in~\autoref{fig:atmsf_analysis}, with the colored area being the 95\% C.L. allowed region, corresponding to $\Delta\chi^2=\chi^2-\chi^2_{\rm min}=5.99$ for 
two-parameter fit.  Different color band corresponds to the $\chi^2$ value.  Four columns in~\autoref{fig:atmsf_analysis} are   $\chi^2_{m_h}$ representing contribution from Higgs mass, $\chi^2_{gg+\gamma\gamma}$ representing contributions from loop induced processes $h\to gg$ and $h\to\gamma\gamma$, $\chi^2_{\mu^\prime}$ representing contributions from tree level Higgs decays to SM fermions and vector bosons, and 
$\chi^2_{total}$, respectively.  Three rows are for $(m_A,\tan\beta)=(1\ {\rm TeV}, 30),\ (2\ {\rm TeV}, 30)$ and $(2\ {\rm TeV}, 3)$, respectively. 

For $m_A=1$ TeV, $\tan \beta=30$, $\chi^2_{m_h}$  (top left panel) could push the $m_{\rm SUSY}$ to be at least 1.4 TeV with $X_t=0$, and larger $m_{\rm SUSY}$ is more preferred for such zero-mixing case.  Two max-mixing branches of $|\frac{X_t}{m_{\rm SUSY}}|\sim 2$ also appear given that the radiative correction to the Higgs mass is the largest under such stop maximal   mixing region.   Larger values of $m_{\rm SUSY}$ are disfavored in these two branches.      $\chi^2_{m_h}$ behaviour for $m_A=2$ TeV, $\tan \beta=30$ (middle left panel) is very similar to that of $m_A=1$ TeV  given that the tree level contributions to $m_h$ are very similar for large $m_A$ at large $\tan\beta$.  Lower left panel of $m_A=2$ TeV, $\tan \beta=3$, however, shows very different behaviour: zero mixing region are completely gone and max-mixing case are preferred with $m_{\rm SUSY}\gtrsim 1.2$ TeV given the need for large radiative corrections with the reduced tree level value of $m_h$.  

For loop induced contributions $\chi^2_{gg+\gamma\gamma}$ (second column),  $m_{\rm SUSY}\le 1$ TeV for zero-mixing case of $X_t=0$ GeV, as well as $|\frac{X_t}{m_{\rm SUSY}}|>3$ are excluded, which corresponds to too large radiative corrections to $h\to gg,\ \gamma\gamma$.  There are, however, large parameter space remains viable in $m_{\rm SUSY}$ vs. $X_t$ plane.  The dependence of $\chi^2_{gg+\gamma\gamma}$ on $m_A$ and $\tan\beta$ is rather weak.

There are strong constraints coming from the precision measurements of Higgs Yukawa and gauge couplings, as shown in $\chi^2_{\mu^\prime}$ plots in the third column.  The most constraining channel is $h\to b\bar{b}$.    As a result, for  $m_A=1$ TeV, $\tan \beta=30$, only positive branch of $X_t$ survives, as shown in the third panel of the top row.  The total $\chi^2_{total}$ include all the contributions gives an even more restricted region of $m_{\rm SUSY}\ge 1.2$ TeV and $X_t/m_{\rm SUSY}\sim 2.6$.   Sensitivity to $\kappa_b$ is reduced for larger values of $m_A$.   For $m_A=2$ TeV, $\tan \beta=30$,  there is larger allowed parameter region when combing all three $\chi^2$s together.   For $m_A=2$ TeV, with small $\tan \beta=3$ (bottom row), while the sensitivity to the Higgs precision measurements are similar to that of the large $\tan\beta$ case, stronger constraints from the Higgs mass lead to the final surviving region  to be  $m_{\rm SUSY}>1.5$ TeV, $|\frac{X_t}{m_{\rm SUSY}}| \approx 2$~\cite{Heinemeyer:2004ms}.

\section{Multi-variable $\chi^2$ fit results}
\label{sec:global}
 
 In this section, we   explore the 95\% C.L. allowed region with the Higgs precision measurements at the CEPC in various MSSM parameter spaces.   With the four most relevant MSSM parameters ($m_A$, $\tan\beta$, $m_{\rm SUSY}$, $X_t$), we scan in the range:
  \begin{eqnarray}\label{eq:para_ana_sec6}
  m_A\in (200, 3000) \text{~GeV}, \tan \beta\in (1, 50),\\ \non
   X_t \in (-5000, 5000) \text{~GeV},  m_{\rm SUSY} \in (200, 3000) \text{~GeV},
\end{eqnarray}
 with $\mu = 500 \text{~GeV}$.   The fitting results vary little when $\mu$ varies.   For the 3D fit performed in our analyses, we fix one variable to a set of benchmark values.  When presenting results in the 2D parameter space, we project the 3D results onto the 2D space for a given set of values of the third parameter.

 \begin{figure}[t]
\begin{center}
\includegraphics[width=5cm]{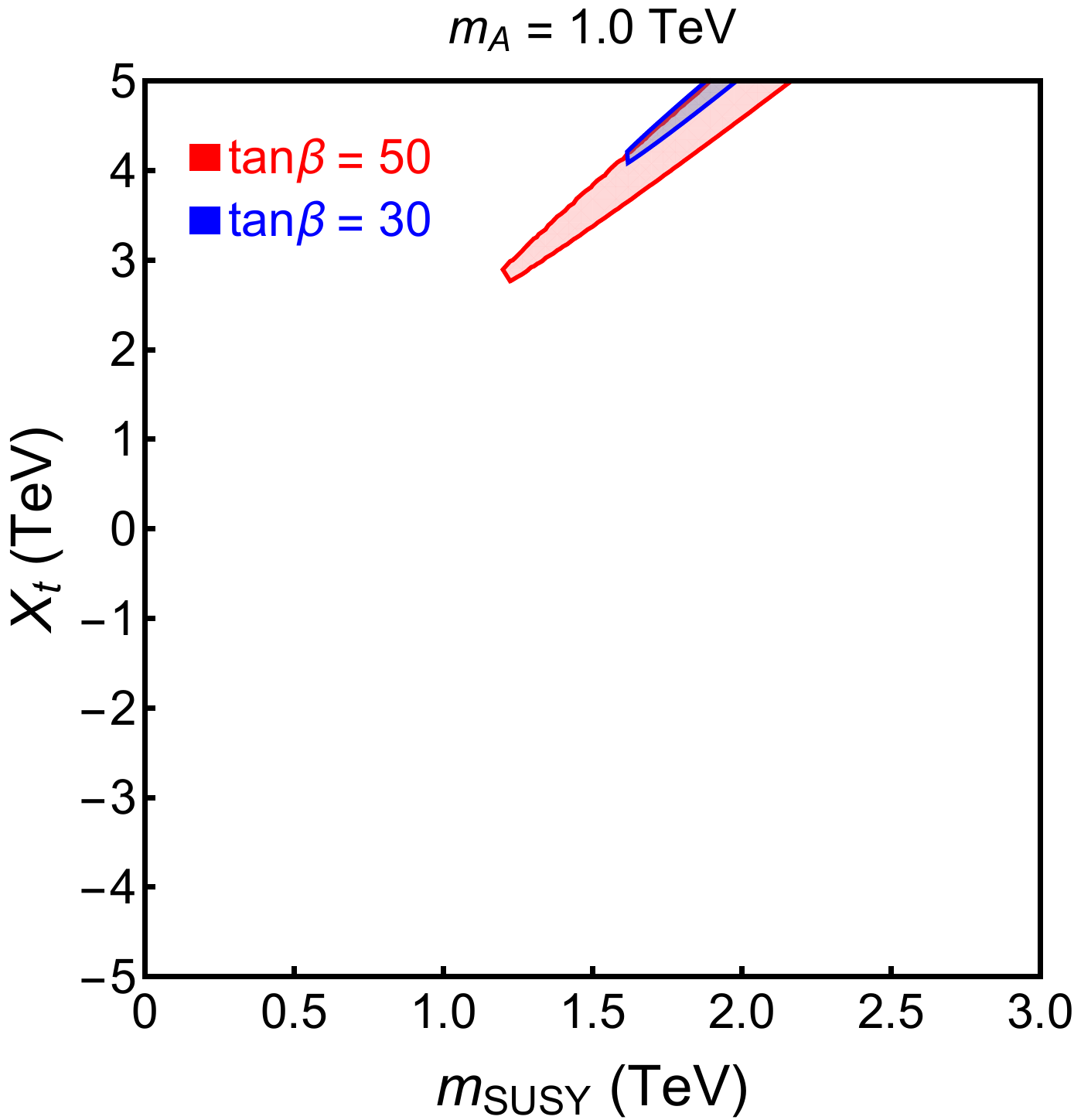}
\includegraphics[width=5cm]{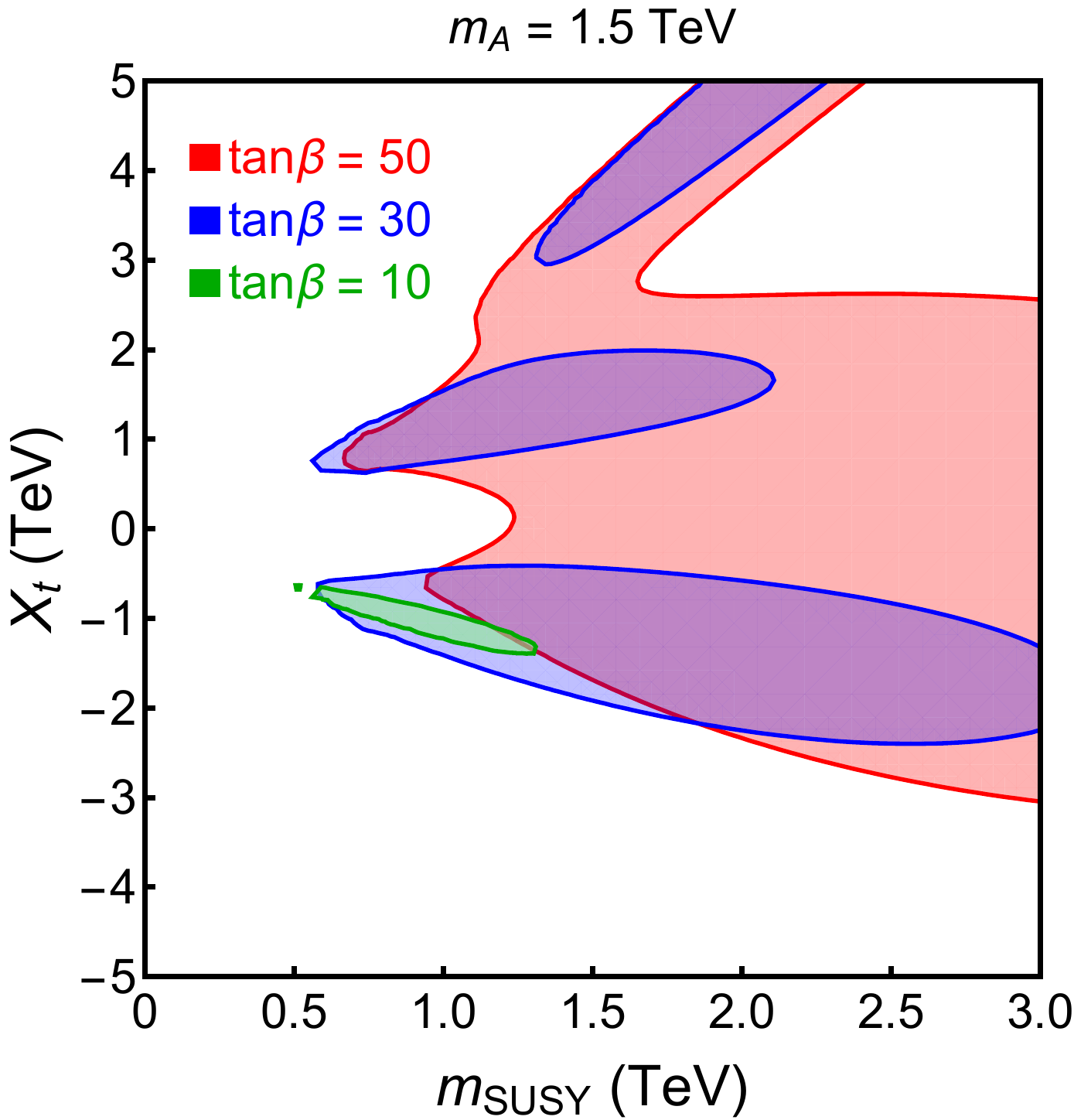}
\includegraphics[width=5cm]{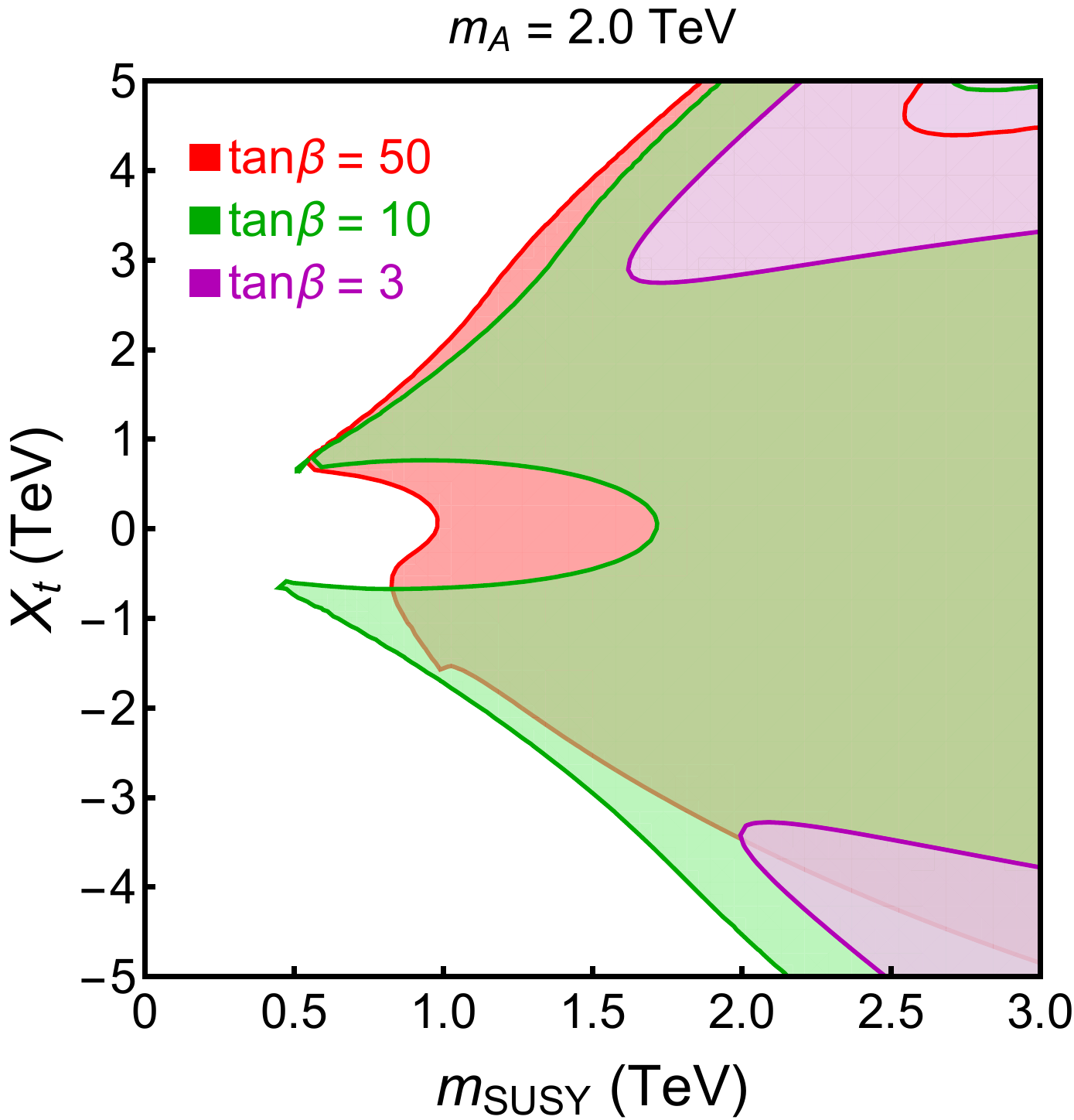}
\caption{95\% C.L. allowed region in $X_t$ vs. $m_{\rm SUSY}$ plane with CEPC precisions for $m_A=1$ TeV (left panel), 1.5 TeV (middle panel)   and 2 TeV (right panel).   For each panel, different colored  region corresponds to different values of $\tan\beta$.
 }
\label{fig:atmsf}
\end{center}
\end{figure}
 
In Fig.~\ref{fig:atmsf}, we show the 95\% C.L. allowed region in $X_t - m_{\rm SUSY}$ plane for various values of $\tan\beta$ with the CEPC precisions.     The left, middle and right panels are for $m_A=1,\ 1.5,\ 2$ TeV, respectively.  The low $\tan \beta$ case receives strong constraints from the Higgs mass precision, especially for smaller values of $m_A$,   as explained in the last section.    For $m_A = 1$ TeV (left panel), $\tan \beta \le 25$ is completely excluded. The survived region is around the stop max-mixing   section of $|X_t|\approx 2 m_{\rm SUSY}$.    Only $X_t>0$ branch survives given the $\kappa_h^b$ effects, as explained in the last section.  For $m_A =1.5$ TeV (middle panel),  $\tan \beta < 10$ is excluded. For $\tan\beta=10$, a small slide of $X_t<0$ survives combining all three contributions to $\chi^2_{total}$.   Larger regions open up for larger values of $\tan\beta$.  For $m_A=2$ TeV (right panel),   $\tan \beta$ as small as 3 is still allowed.    Precision constraints from both the mass and the couplings  are relaxed for larger $\tan\beta$ and larger $m_A$, resulting in large survival parameter spaces in $X_t$ vs. $m_{\rm SUSY}$.  
\begin{figure}[htb]
\begin{center}
\includegraphics[width=6.5cm]{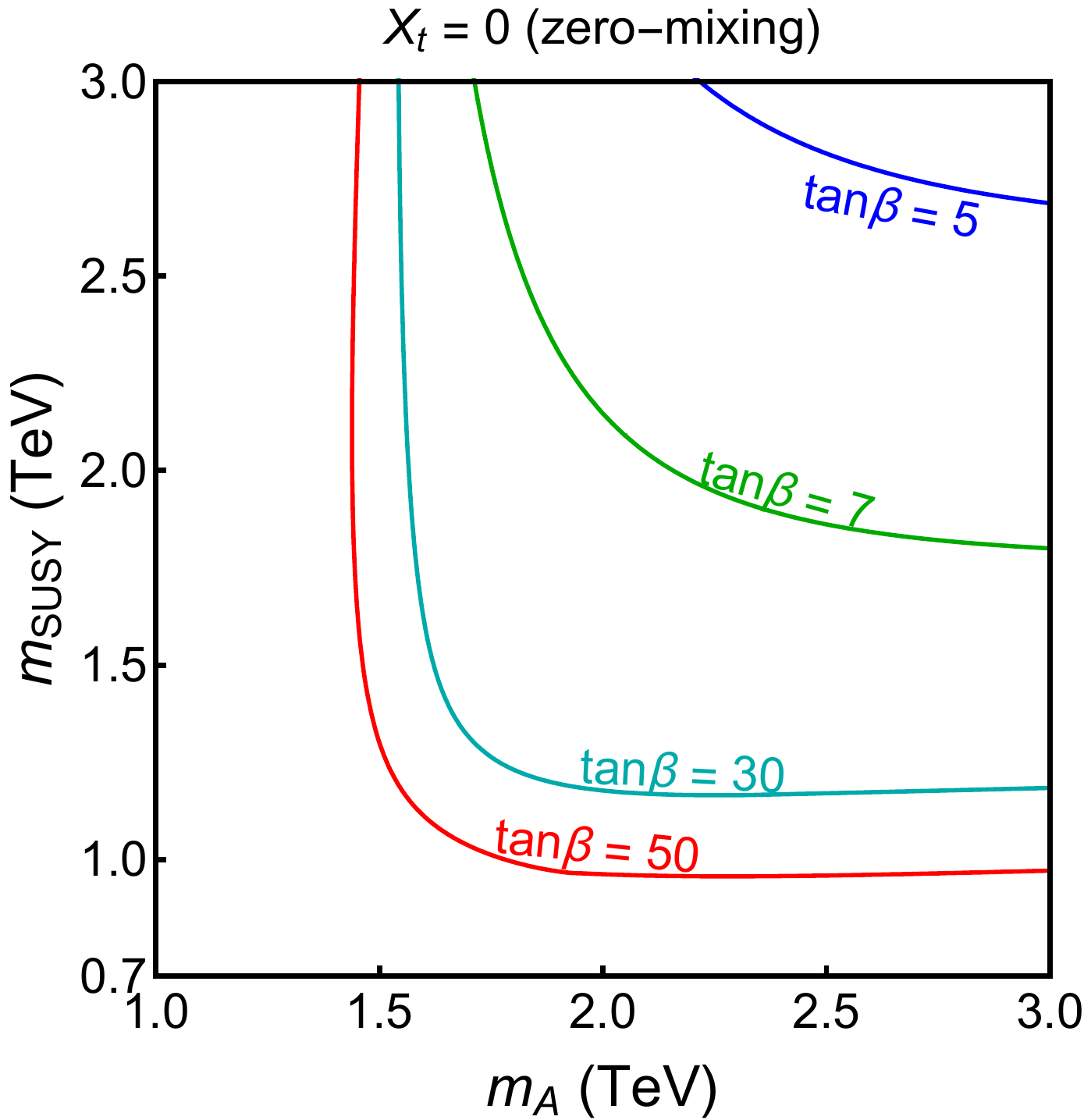}
\includegraphics[width=6.5cm]{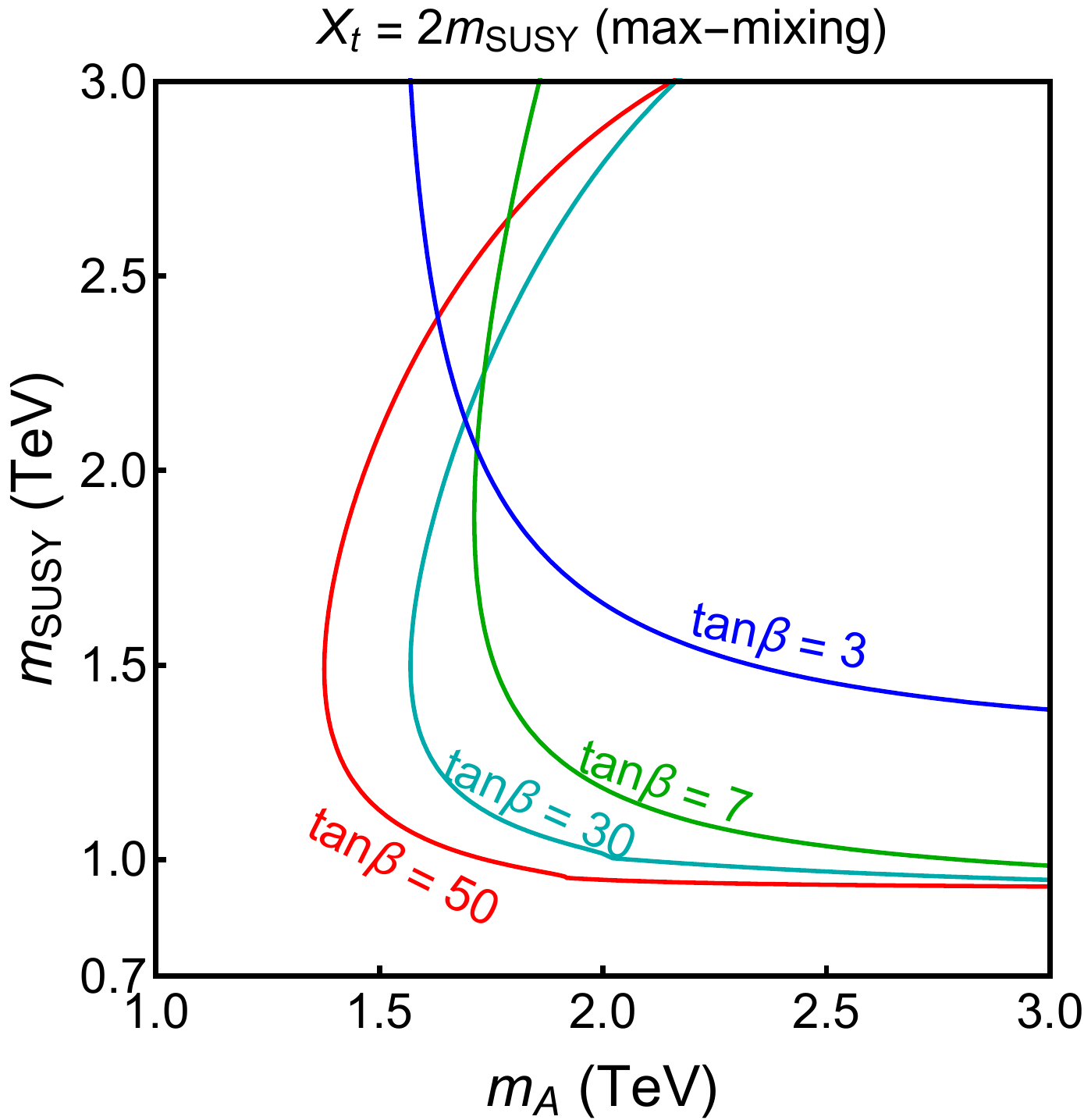}
\caption{95\% C.L. allowed region in $m_{\rm SUSY}$ vs. $m_A$ plane with CEPC precisions for $X_t =0$ (zero-mixing, left panel)  and $X_t = 2 m_{\rm SUSY}$ (max-mixing, right panel).   For each panel, different colored curve corresponds to different values of $\tan\beta$, with region to the right of the curve allowed.   
}
\label{fig:mamsf}
\end{center}
\end{figure}

In Fig.~\ref{fig:mamsf}, we show the 95\% C.L. allowed region in $m_{\rm SUSY}$ vs. $m_A$ plane with the CEPC precisions for $X_t =0$ (zero-mixing, left panel)  and $X_t = 2 m_{\rm SUSY}$ (max-mixing, right panel).  Regions to the right of the curve are the 95\% C.L. allowed regions for different values of $\tan\beta$.    For the zero-mixing case,  small $\tan\beta$ receives the strongest constraints, with $\tan \beta \leq 4 $ is excluded totally. For  $\tan \beta = 50$,   $m_A\ge 1350$ GeV and $m_{\rm SUSY}\ge 850$ GeV are still allowed.   Note that for the zero-mixing case, the most important constraints for $m_A$ come from the Higgs gauge and Yukawa couplings, while the most important constraint  for   $m_{\rm SUSY}$ comes from Higgs mass precision except for the large $\tan \beta$ case, when the Higgs gauge and Yukawa couplings enter as well.

In the max-mixing case, value of $\tan\beta$ as low as 3 could be accommodated.  The allowed region is typically larger comparing to that of the zero-mixing case.   For small $\tan\beta$, the strongest constraints for $m_{\rm SUSY}$  are Higgs mass precision and loop-induce $hgg$ and $h\gamma\gamma$.  For $\tan \beta \geq 7$, the lower limit on $m_{\rm SUSY}$ mostly comes from the loop-induce $hgg$ and $h\gamma\gamma$, which is less sensitive to values of $\tan\beta$.  There are, however, upper limit on $m_{\rm SUSY}$ comes from too large contributions to $m_h$.   Limits on $m_A$ are mostly determined by the precisions of Higgs couplings.  

\begin{figure}[t]
\begin{center}
\includegraphics[width=7.5cm]{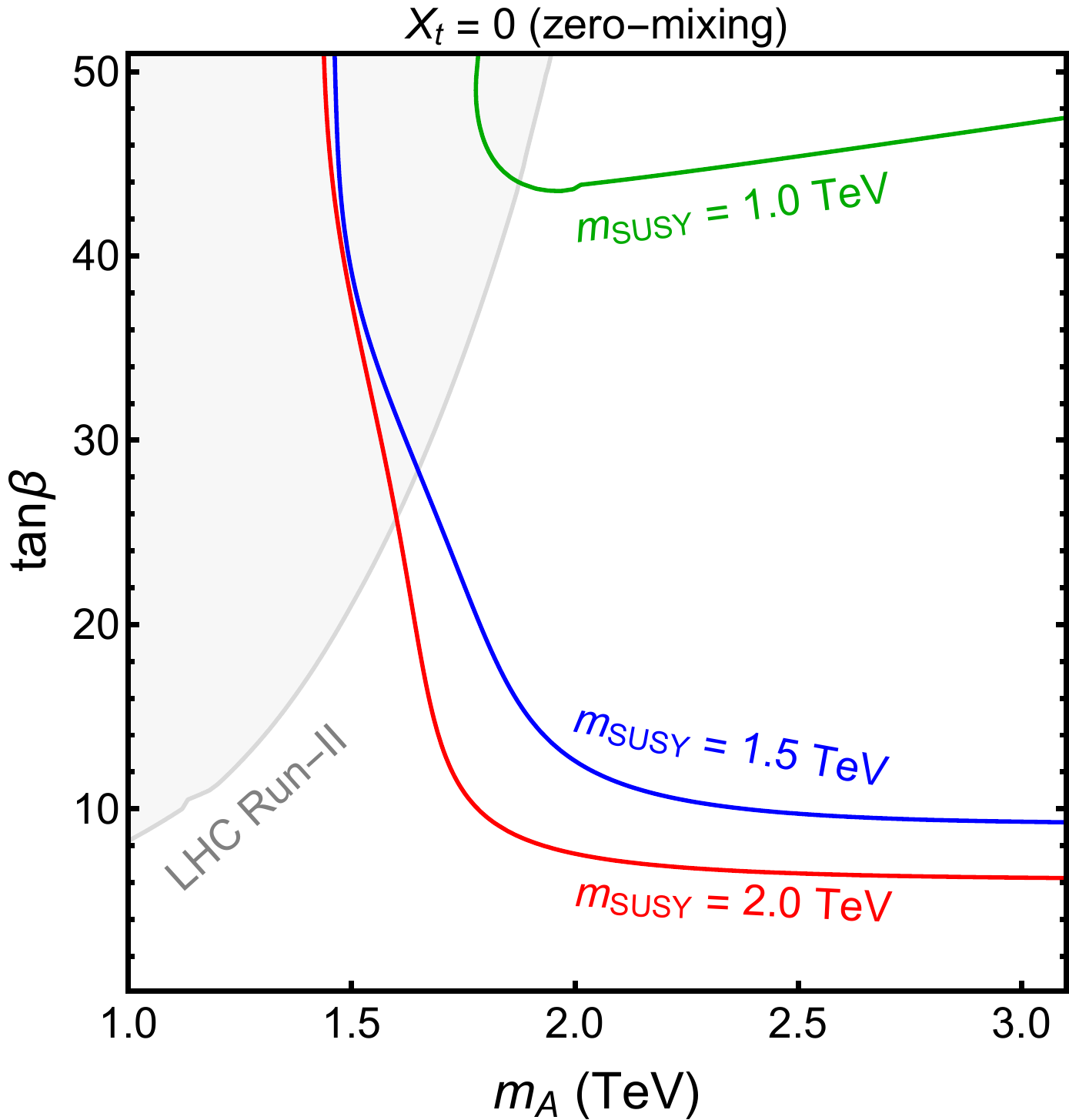}
\includegraphics[width=7.5cm]{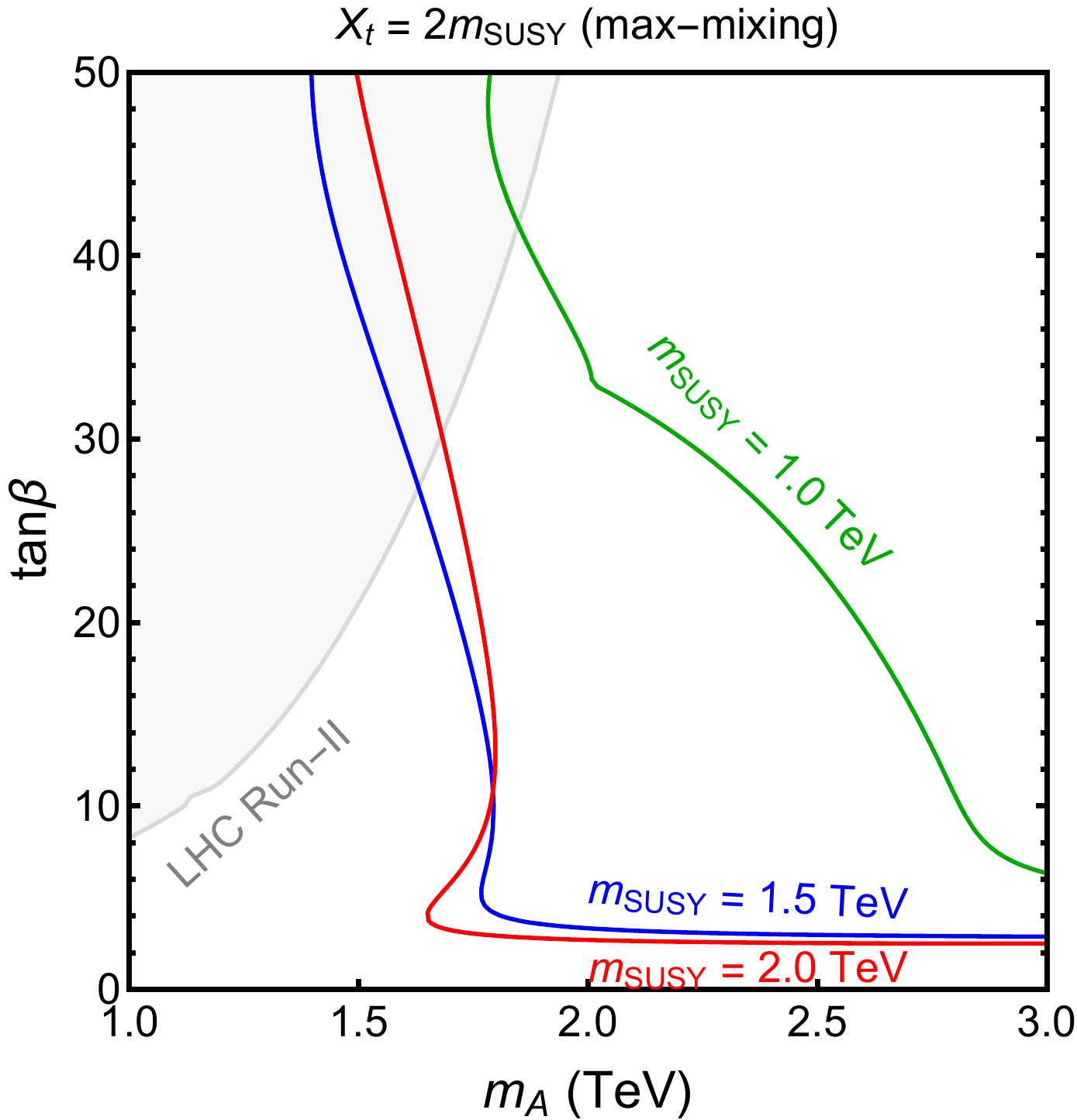}
 \caption{ 95\% C.L. allowed region in $\tan\beta$ vs. $m_A$ plane with the CEPC precisions for $X_t =0$ (zero mixing, left panel)  and $X_t = 2 m_{\rm SUSY}$ (max-mixing, right panel).   For each panel, different colored curve corresponds to different values of $m_{\rm SUSY}$, with region above the curve allowed.  The LHC Run-II direct search limits based on $A/H\to\tau\tau$~\cite{Aad:2020zxo} are shown in the grey shaded region.  }
\label{fig:matanb}
\end{center}
\end{figure}

The results of the three-parameter fit for $m_A, m_{\rm SUSY}$ and $\tan\beta$ are projected onto $m_A$ vs. $\tan\beta$ plane in Fig.~\ref{fig:matanb}. Regions above the curve are the 95\% C.L. surviving regions with   CEPC precisions for different values of $m_{\rm SUSY}$.   In general,  $m_{\rm SUSY}<900$ GeV are excluded for both the no-mixing and max-mixing cases.  For the no-mixing case, when $m_{\rm SUSY}<1$ TeV, $\tan\beta<40$ is excluded. Limits on $\tan\beta$ get lower for larger values of $m_{\rm SUSY}$, which is sensitive in particular for $1\ {\rm TeV}<m_{\rm SUSY}<1.5$ TeV. For the max-mixing case, limits on $\tan\beta$ is much lower for $m_{\rm SUSY}=1$ TeV.    Those features   are mainly due to the Higgs mass constraint.  The LHC Run-II direct search limits based on $A/H\to\tau\tau$~\cite{Aad:2020zxo} are shown in the grey shaded region, which is complementary to the indirect limits from Higgs precision measurements.

\begin{figure}[t]
\begin{center}
\includegraphics[width=7.5cm]{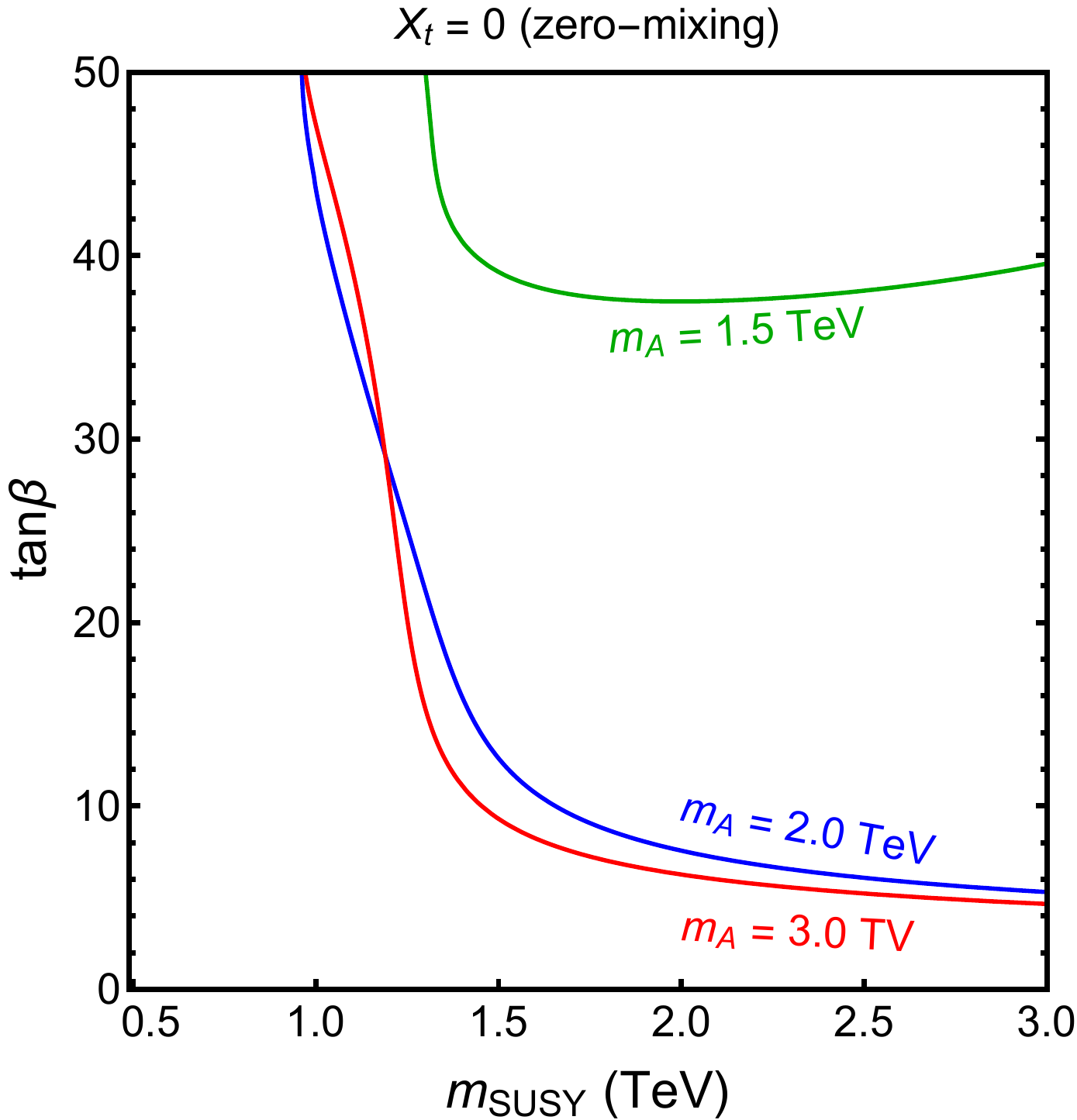}
\includegraphics[width=7.5cm]{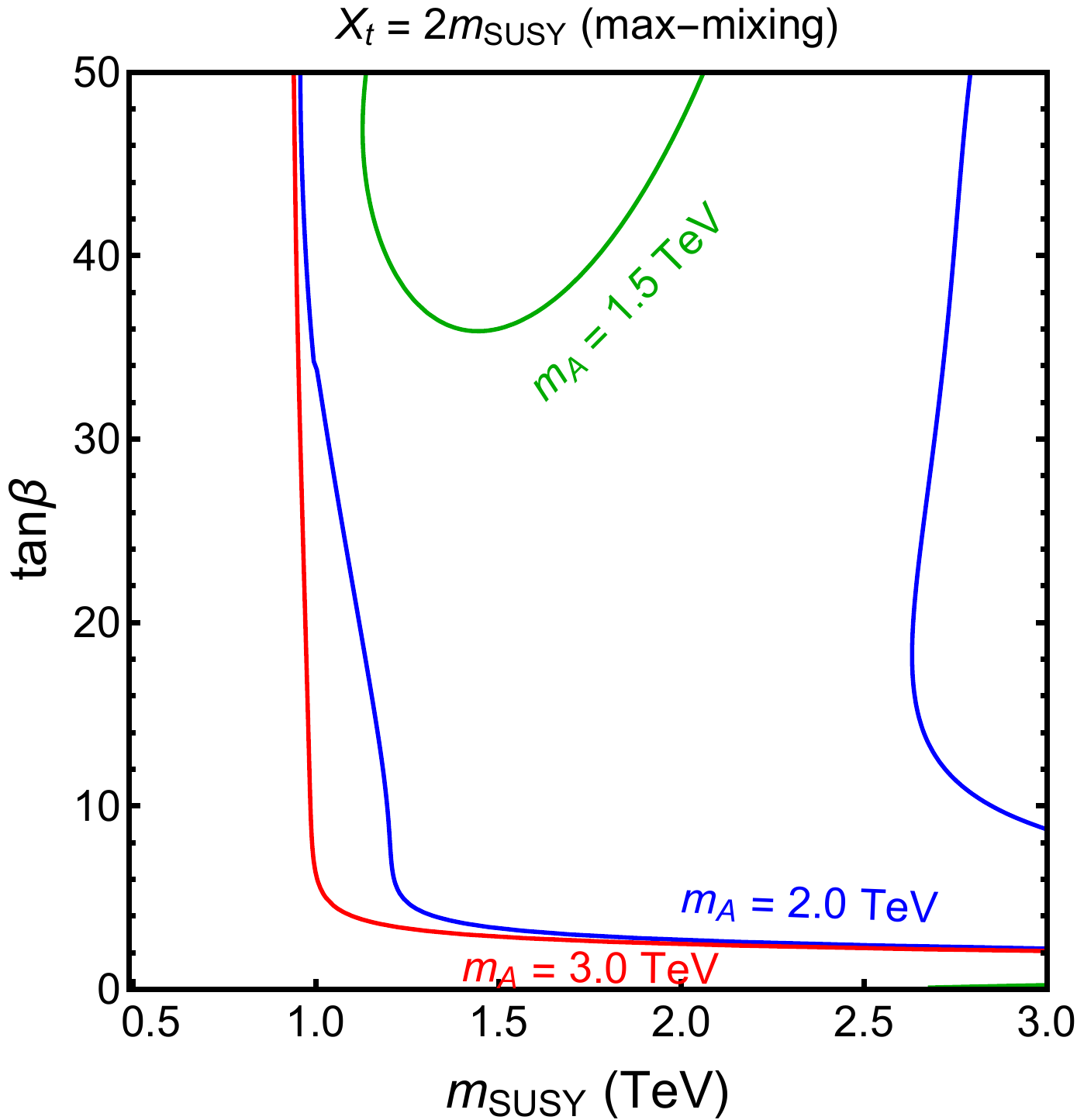}
\caption{ 95\% C.L. allowed region in $\tan\beta$ vs. $m_{\rm SUSY}$ plane with the CEPC precisions for $X_t =0$ (zero-mixing, left panel)  and $X_t = 2 m_{\rm SUSY}$ (max-mixing, right panel).   For each panel, different colored curve corresponds to different values of $m_A$, with region above the curve allowed, except for the $m_A=2$ TeV (blue curves) in the right panel, in which region between two curves is allowed.
}
\label{fig:msusytanb}
\end{center}
\end{figure}

The results of the three-parameter fit for $m_A, m_{\rm SUSY}$ and $\tan\beta$ are projected onto $m_{\rm SUSY}$ vs. $\tan\beta$ plane in Fig.~\ref{fig:msusytanb}. For each panel, different colored curve corresponds to different values of $m_A$, with region above the curve allowed, except for the $m_A=2$ TeV (blue curves) in the right panel, in which region between two curves is allowed.  In general,  $m_A<1$ TeV are excluded for both the zero-mixing and max-mixing cases.   The lower limits on $\tan\beta$ are relaxed for larger values of $m_A$, and is sensitive to the values of $m_A$  for $1.5\ {\rm TeV}<m_A<2$ TeV.   For the max-mixing case and a given $m_A$,  there are the upper limit for $m_{\rm SUSY}$ at large $\tan\beta$, as shown in the right panel of Fig.~\ref{fig:msusytanb}.   This is due to the too large contribution to $m_h$ for larger values of $m_{\rm SUSY}$.  
 For $m_A=3$ TeV, the upper limit for $m_{\rm SUSY}$ is larger than 3 TeV, therefore not shown in the plot.

\begin{figure}[htb]
\begin{center}
\includegraphics[width=6.5cm]{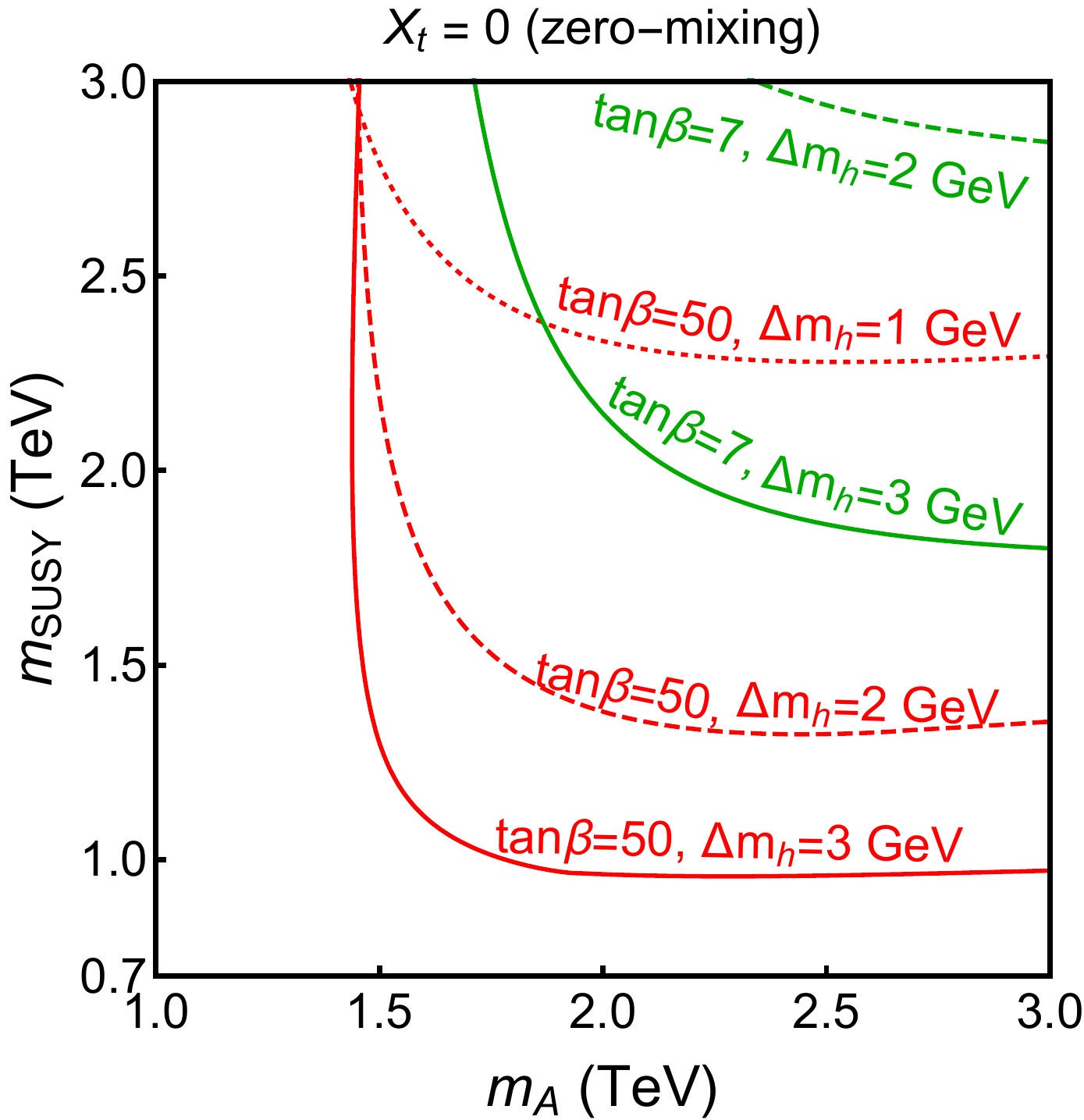}
\includegraphics[width=6.5cm]{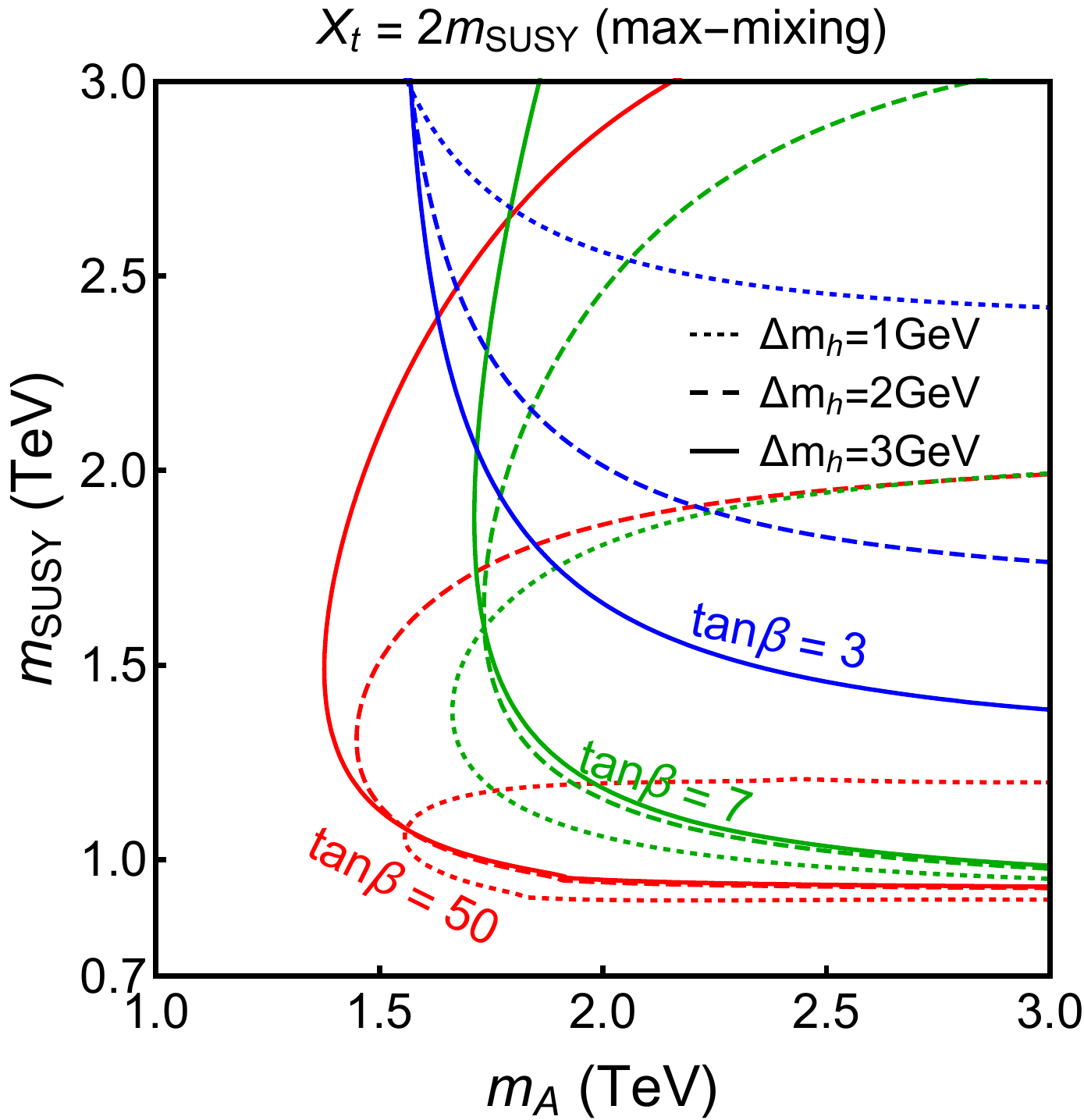}
\caption{95\% C.L. allowed region in $m_{\rm SUSY}$ vs. $m_A$ plane with CEPC precisions for $X_t =0$ (zero-mixing, left panel)  and $X_t = 2 m_{\rm SUSY}$ (max-mixing, right panel).   For each panel, different colored curve corresponds to different values of $\Delta m_h=1,\ 2, \ 3$ GeV, with region to the right of the curve allowed. 
}
\label{fig:deltamh1GeV}
\end{center}
\end{figure}

To illustrate the potential impact of future improvement in the MSSM prediction of $m_h$,  in Fig.~\ref{fig:deltamh1GeV}, we show the  95\% C.L. allowed region in $m_A$ vs. $m_{SUSY}$ plane for $\Delta m_h=3$ GeV (solid curve) 2 GeV (dashed curve) and 1 GeV (dotted curve).      The lower limit on $m_{\rm SUSY}$ for the zero-mixing case, and the upper limit on $m_{\rm SUSY}$ for the max-mixing case   depend sensitively on the values of $\Delta m_h$.    Therefore, it is crucial to improve the precision in the $m_h$ calculation in the MSSM, which allows us to obtain tight constraints on the SUSY mass scale, in particular, on the stop sector, once Higgs precision measurements are available at future Higgs factories.

\section{Comparison between different Higgs factories}
\label{sec:comparison}

\begin{figure}[htb]
\begin{center}
\includegraphics[width=6.5cm]{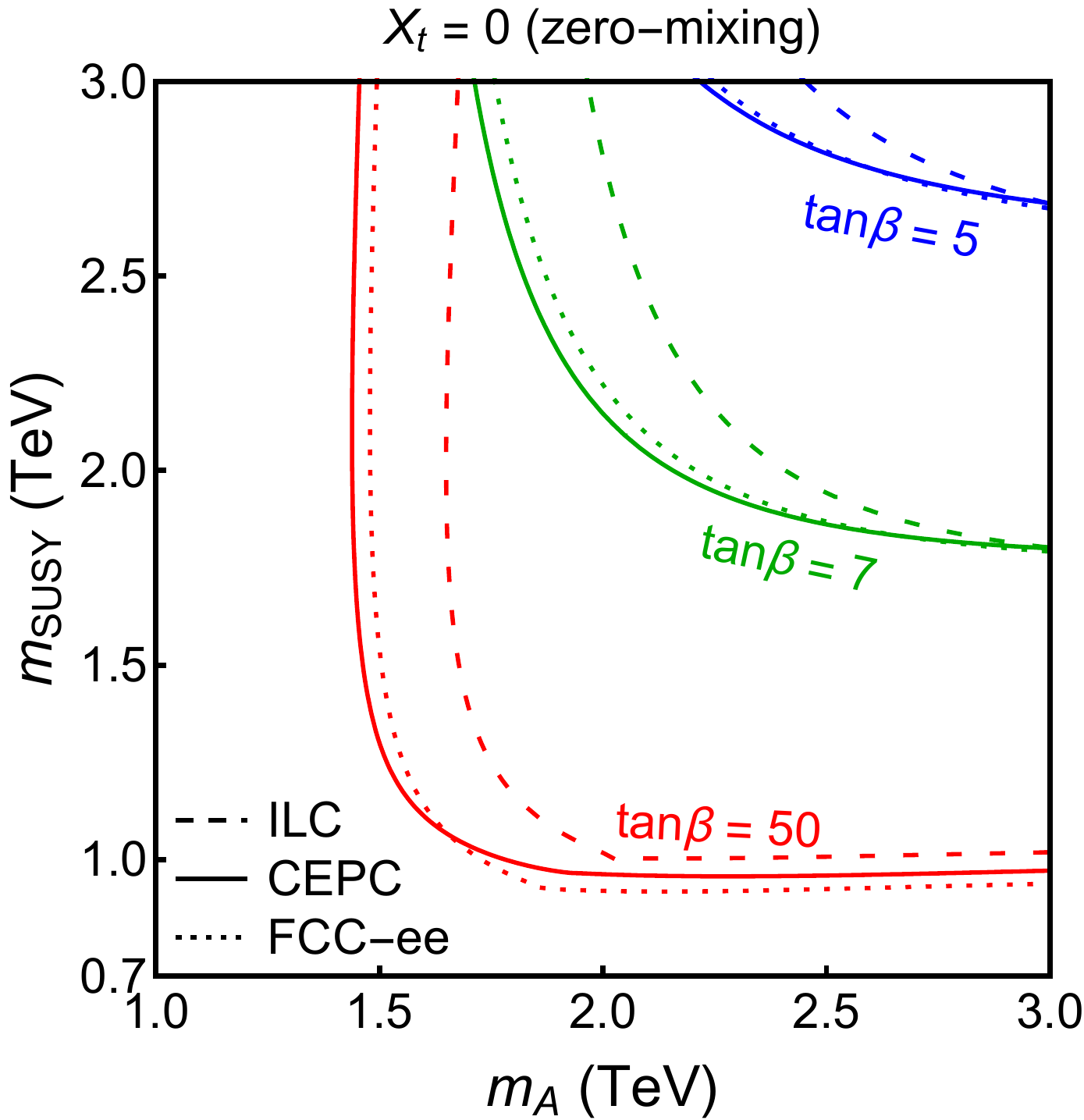}
 \includegraphics[width=6.5cm]{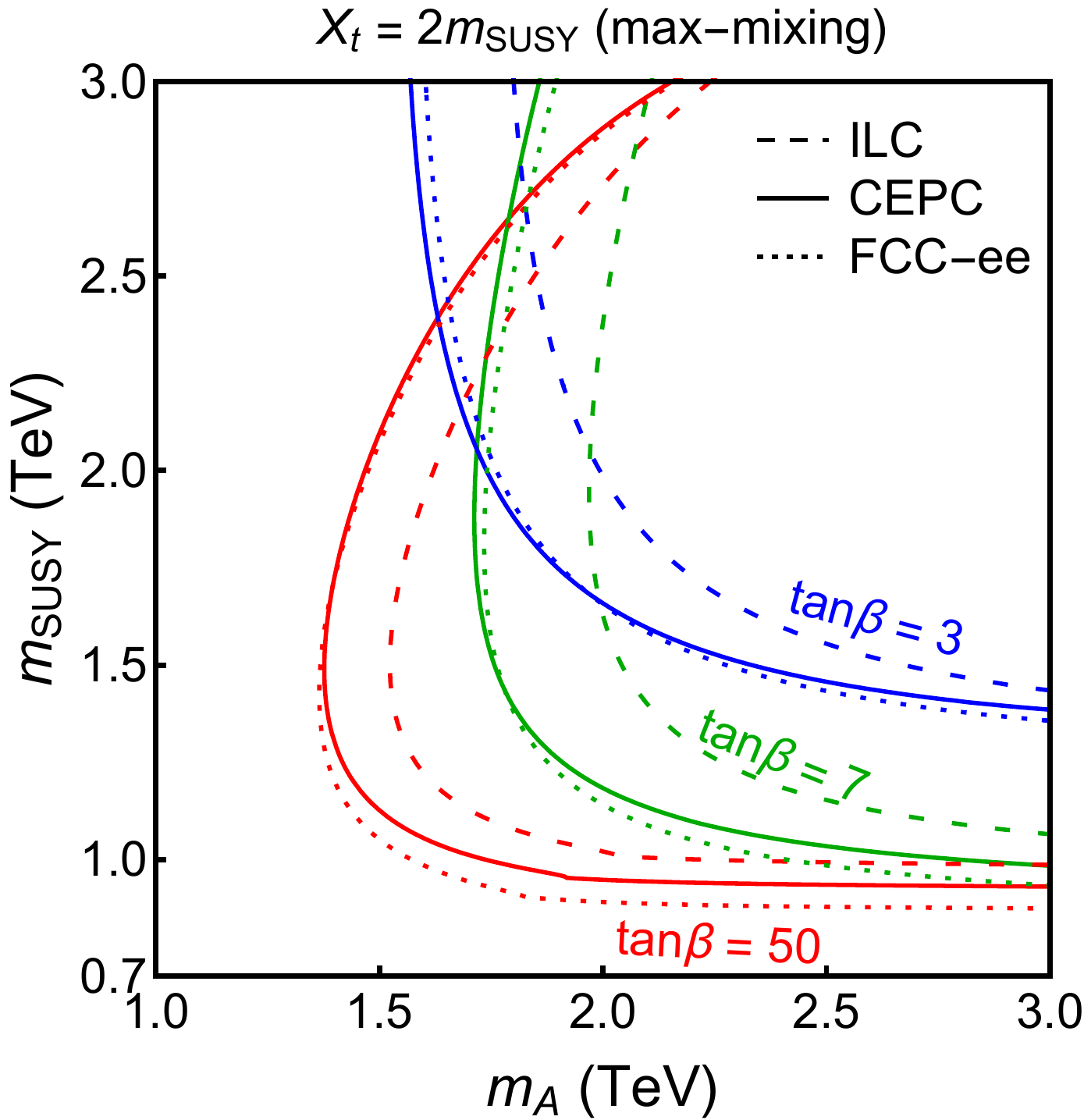}
\includegraphics[width=6.5cm]{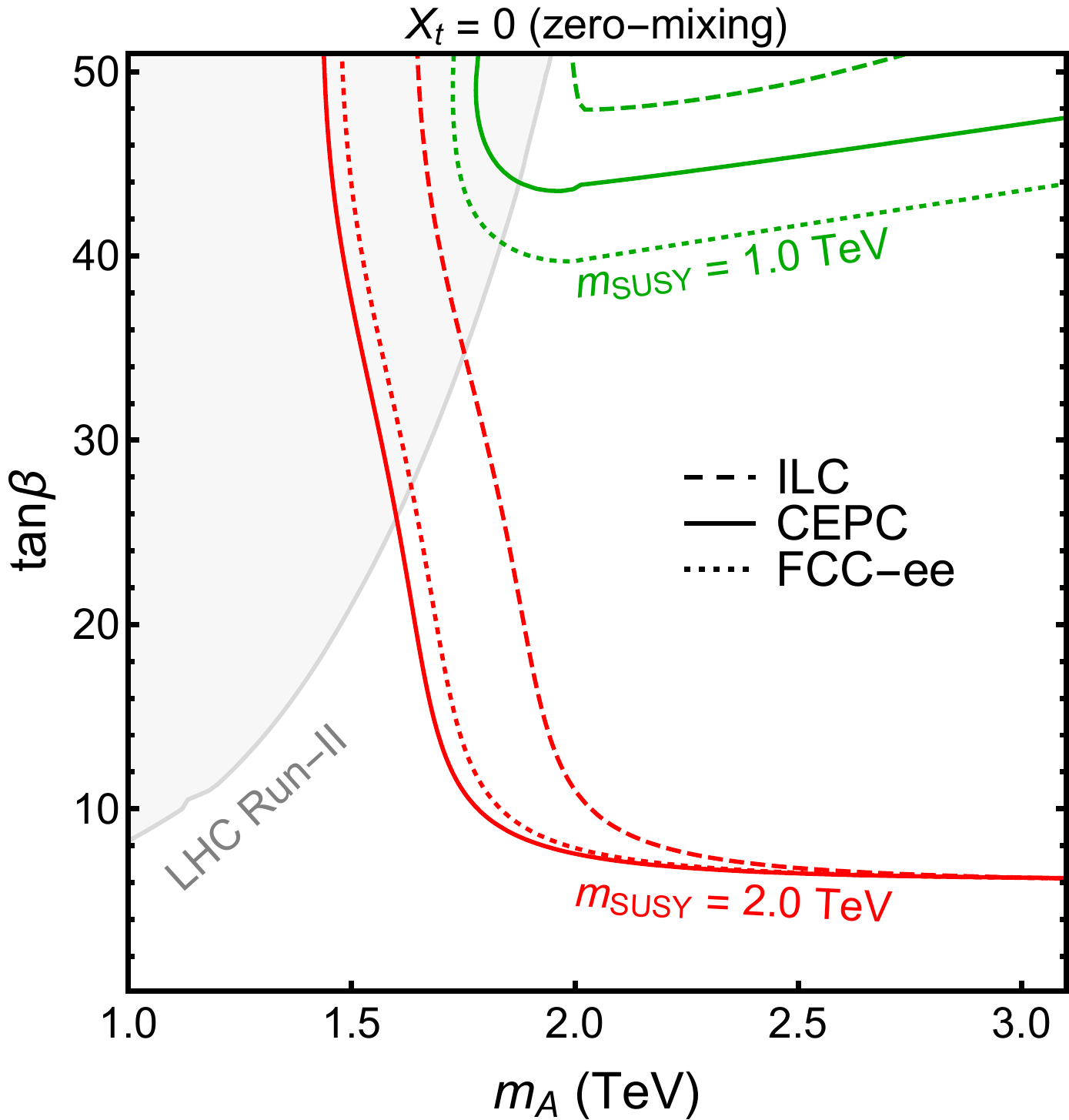}
\includegraphics[width=6.5cm]{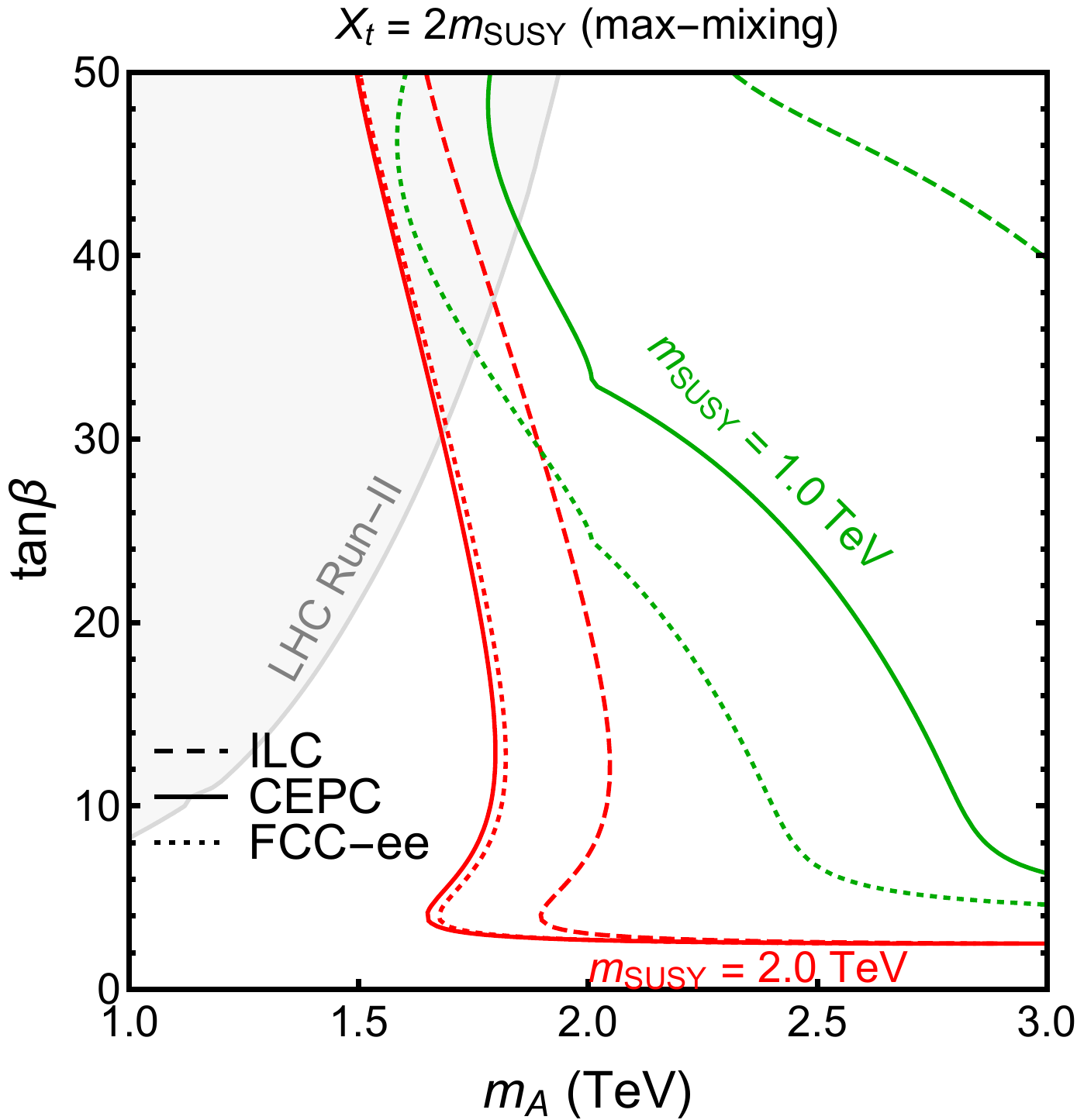}
 \caption{95\% C.L. allowed region in $m_{\rm SUSY}$ vs. $m_A$ plane (top row), and $\tan\beta$ vs. $m_A$ (bottom row)  for $X_t =0$ (zero-mixing, left panels)  and $X_t = 2 m_{\rm SUSY}$ (max-mixing, right panels), with  the CEPC (solid), the FCC-ee (dotted) and the ILC (dashed) precision.    The LHC Run-II direct search limits based on $A/H\to\tau\tau$~\cite{Aad:2020zxo} are shown in the grey shaded region in the bottom panels.
 }
\label{fig:comparison}
\end{center}
\end{figure}
 
To compare the reach for three different Higgs factories, we show the 95\% C.L. allowed region in Fig.~\ref{fig:comparison} in the parameter space of $m_{\rm SUSY}$ vs. $m_A$ (upper two panels) and $\tan\beta$ vs. $m_A$ (lower two panels) for the zero-mixing and max-mixing cases.  While the CEPC and the FCC-ee has similar reach, the reach in $m_A$ for the ILC is better because of better precisions on Higgs measurements of $hbb$ and $hWW$, given the increased rate of the WBF processes at higher center of mass energies.  Limits of $m_{\rm SUSY}$ and $\tan\beta$ (for large $m_{\rm SUSY}$)  are nearly the same for all the three Higgs factories because it is mainly controlled by the precision in Higgs mass which   comes from theoretical uncertainties.     Note that for the $m_A$ vs. $\tan\beta$ plot, the CEPC results are more constraining than the FCC-ee results for $m_{\rm SUSY}=1$ TeV, while nearly the same for $m_{\rm SUSY}=2$ TeV.  This is due to the slightly better precision in $h\to gg, \gamma\gamma$ channel at the CEPC, 
which makes it more sensitive for smaller stop mass running in the loop.  

\section{Conclusion and outlook}
\label{sec:conclusion}

In this work, we studied the constraints of Higgs precision measurements from future Higgs factories on the MSSM parameter space.  We considered the dominant stop contributions to the Higgs mass, loop induced $h\gamma\gamma+hgg$ couplings, effective mixing angle $\alpha_{eff}$, which enters the Higgs couplings to pair of fermions and gauge bosons,  as well as additional loop contributions to the bottom Yukawa coupling $\kappa_b$.  The four relevant parameters under consideration are  $m_A$, $\tan\beta$, $m_{\rm SUSY}$ and $X_t$.

 In the multi-variable $\chi^2$ fit, we included all the Higgs decay channels to SM fermions and gauge bosons at Higgs factories, as well as the Higgs mass.   We found that $\chi^2_{m_h}$ dominates for the small $\tan\beta$ case, while $\chi^2$ contributions from   the Higgs decays, in particular, $h\rightarrow b\bar{b}$, dominates for the small to moderate $m_A$ case.  Generally we found CP-odd Higgs mass $m_A$ is sensitive to the precisions of Higgs decay channels,  while $m_{\rm SUSY}$, $X_t$ and $\tan\beta$ are sensitive to the precision of Higgs mass determination. For large $\tan\beta$,  $m_{\rm SUSY}$ and $X_t$ are also sensitive  the precisions of fermion and vector gauge boson couplings. For the max-mixing scenario, the loop-induced $hgg$ and $h\gamma\gamma$ couplings are the main restrictions on $m_{\rm SUSY}$ when $\tan\beta>7$.

 We obtained the 95\% C.L. allowed region given the Higgs factory precisions, and presented the result in the parameter space of $M_{\rm SUSY}$ vs. $X_t$, $M_A$ vs. $m_{\rm SUSY}$, $m_A$ vs. $\tan\beta$ and $m_{\rm SUSY}$ vs. $\tan\beta$.  We found that small $\tan\beta$ only survives in the max-mixing case with relatively large $m_A$, while large regions of $X_t$  vs. $m_{\rm SUSY}$ are allowed for large $\tan\beta$ and large $m_A$.   The lower limits on $\tan\beta$ depends sensitively on the values of $m_{\rm SUSY}$ and $m_A$, in particular, for $m_{\rm SUSY}<1.5$ TeV and $m_A<2$ TeV.   Limits on $m_{\rm SUSY}$ also depend sensitively on $\Delta m_h$, indicating the importance of a precise determination of the Higgs mass in the MSSM. For $\tan\beta=50$ of the max-mixing scenario, $m_{\rm SUSY}  \in (0.8,1.2) \gev$ when $\Delta m_h= 1 $ GeV. 
 
We also compared the reach of the CEPC, the FCC-ee and the ILC.  We found that the reach of the CEPC is similar to that of the FCC-ee, while the reach of the ILC is typically better, given the slight better precision in the Higgs WBF measurements.   With the high precision of the Higgs coupling measurements, and the potential improvement of theoretical calculation of $m_h$ in the MSSM, studying the SM-like Higgs properties at future Higgs factories offer great insight into the MSSM parameter space, which will be complementary to the direct searches of SUSY particles at   energy frontier machines.

\begin{acknowledgments}
We thank S. Heinemeyer for insightful discussions on the Higgs sector of the MSSM. HL was supported  by  the National Natural Science
Foundation of China (NNSFC) under grant No. \texttt{11635009} and Natural Science Foundation of Shandong Province under grant No. \texttt{ZR2017JL006}.
HS and SS are supported in part by the Department of Energy under Grant \texttt{DE-FG02-13ER41976/DE-SC0009913}.  
WS is supported by the Australian Research Council (ARC) Centre of Excellence for Dark Matter Particle Physics (CE200100008).
JMY  was supported by the National Natural Science Foundation of China (NNSFC) under grant Nos.11675242, 12075300, 11821505, and 11851303, by Peng-Huan-Wu Theoretical Physics Innovation Center (11947302), by the CAS Center for Excellence in Particle Physics (CCEPP), by the CAS Key Research Program of Frontier Sciences and by a Key RD Program of Ministry of Science and Technology under number 2017YFA0402204.
\end{acknowledgments}


\bibliographystyle{JHEP}
\bibliography{mssm}

\end{document}